\pdfoutput=1
\documentclass[PAPER, atlasdraft=false, UKenglish, cernpreprint, texmf, orcidlogo]{atlasdoc}
\usepackage{atlaspackage}
\usepackage{atlasbiblatex}

\usepackage{atlasphysics}
\graphicspath{{logos/}{figures/}}
\usepackage{multirow}

\usepackage{rotating}

\usepackage{ANA-EXOT-2022-44-PAPER-defs}

\AtlasTitle{Search for a top-philic heavy resonance in association with top quarks in $pp$ collisions at $\sqrt{s} =$ 13 TeV and 13.6 TeV with the ATLAS detector}

\AtlasAbstract{%
A search for a new top-philic vector boson ($Z'$) produced in association with
a top-quark or a pair of top quarks, and decaying into a top quark pair is presented.
The search uses $pp$ collision data collected by the ATLAS detector at the Large
Hadron Collider at centre-of-mass energies of $\sqrt{s} =$ 13 TeV and 13.6 TeV,
corresponding to integrated luminosities of 140~fb$^{-1}$ and 56~fb$^{-1}$, respectively.
Events containing exactly two same-sign leptons or at least three leptons (electrons or muons) are selected.
A transformer-based classifier parameterised as a function of the $Z'$ mass is used to optimise
the signal-to-background discrimination.
The results are interpreted in a simplified model with a colour-singlet vector boson
coupled to top quarks with coupling strength $c_t$ and chirality angle $\theta$.
No significant excess is observed above the expected Standard Model background.
A $Z'$ boson with mass below 1150 GeV is excluded at 95\% confidence
level for a benchmark scenario with $c_t = 1$ and $\theta = 0$, with
an expected limit of 1250 GeV.
}

\AtlasRefCode{EXOT-2022-44}

\PreprintIdNumber{CERN-EP-2026-249}

\AtlasJournal{JHEP}
\AtlasCoverEgroupAnalysisTeam{atlas-exot-2022-44-analysis-team@cern.ch}

\hypersetup{pdftitle={ATLAS document},pdfauthor={The ATLAS Collaboration}}

\begin{document}

\maketitle

\tableofcontents

\section{Introduction}
%

% The next lines are included from the .//sections/1-introduction.tex input file
The top quark, discovered in 1995 by the CDF and D0 Collaborations~\cite{CDF:1995wbb,D0:1995jca},
is the heaviest elementary particle in the Standard Model (SM) of particle physics.
With a mass close to the electroweak symmetry-breaking scale, the top
quark plays an important role in the dynamics responsible for electroweak symmetry breaking
and in many possible extensions of the SM\@.
Precision measurements of top-quark production and decay properties
have so far shown good agreement with the SM predictions, as reviewed in Ref.~\cite{TOPQ-2023-19}, establishing the SM as an effective description of nature up to the \TeV~energy scale.
However, several important theoretical questions remain unresolved. In particular,
the stability of the electroweak scale in the presence of physics at much higher energy
scales remains a central motivation for many extensions of the SM~\cite{Susskind:1978ms}.
Owing to its large Yukawa coupling, processes involving top quarks
are therefore especially sensitive to physics beyond the Standard Model (BSM).

As a result, many BSM theories addressing these issues, including composite Higgs
models~\cite{Ferretti:2013kya,Vecchi:2015fma,Agashe:2003zs,Agashe:2004rs}, introduce new particles
associated with the dynamics responsible for stabilizing the electroweak scale.
These states consequently couple preferentially to the top sector.
Characterised by suppressed couplings to light quarks and gluons,
their production at the Large Hadron Collider (LHC) is expected
to proceed predominantly through interactions involving top quarks.
This naturally leads to experimental signatures with multiple top quarks in the final state,
including events containing three or four top quarks.
Such signatures were investigated in measurements of SM four-top-quarks (\tttt) production in ATLAS~\cite{TOPQ-2021-08-witherratum},
where a cross-section of $22.5^{+6.6}_{-5.5}$ fb was measured, compared with the SM prediction of $12.0 \pm 2.4$ fb,
further motivating dedicated searches for top-philic resonances in multi-top-quark final states.

\begin{figure}[!htbp]
\centering
\begin{minipage}[b]{0.28\linewidth}
\centering
\includegraphics[width=\linewidth]{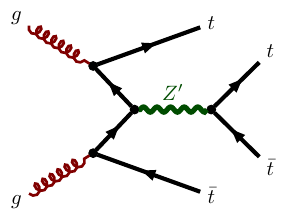}
\par (a)
\end{minipage}
\hspace{0.0\linewidth}
\begin{minipage}[b]{0.20\linewidth}
\centering
\includegraphics[width=\linewidth]{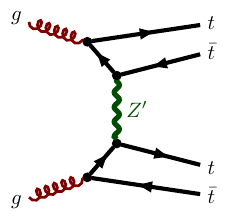}
\par (b)
\end{minipage}
\hspace{0.0\linewidth}
\begin{minipage}[b]{0.26\linewidth}
\centering
\includegraphics[width=\linewidth]{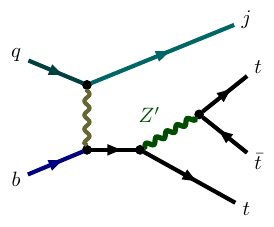}
\par (c)
\end{minipage}
\hspace{0.0\linewidth}
\begin{minipage}[b]{0.22\linewidth}
\centering
\includegraphics[width=\linewidth]{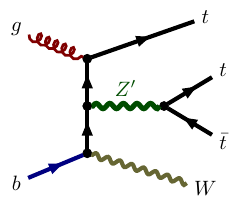}
\par (d)
\end{minipage}
\caption{\label{fig:feyn} Representative Feynman diagrams for the production of a top-philic \Zp resonance in association with (a, b) a top--anti-top pair or (c, d) a single top quark. The four-top diagrams illustrate \Zp exchange in the (a) $s$-channel and (b) $t$-channel, while the three-top diagrams show \Zp production in association with a top quark and (c) a light quark, $tjZ^\prime$, or (d) a $W$ boson, $tWZ^\prime$ (denoted collectively \tZp).}
\end{figure}

This paper presents a search for a new top-philic heavy resonance with a mass above 400~\GeV, produced in association with one or two top quarks.
Representative Feynman diagrams for these processes are shown in Figure~\ref{fig:feyn}.
A simplified model~\cite{Kim_2016} is used to generate simulated signal samples and to interpret the results in a model-dependent framework.
In this framework, the new particle is described as a colour-singlet vector boson, denoted \Zp.
A direct search for \Zp bosons produced in association with two top quarks was performed
by the ATLAS Collaboration in single-lepton final state using full Run~2 data, constraining the production
cross-section of a \Zp boson coupling only to top quarks to values ranging from 21 to 119~fb
over the mass range 1--3~\TeV~\cite{EXOT-2022-14}.
Similar multi-top-quark final states have also been explored by the ATLAS and CMS
Collaborations in searches for top-philic heavy resonances in three-top- and four-top-quark events~\cite{CMS-TOP-24-008},
as well as heavy scalar and pseudo-scalar Higgs bosons produced in association with a top-quark pair~\cite{EXOT-2019-26-witherratum}.
Allowing for couplings to light quarks but not leptons, searches for $\Zp \rightarrow t\bar{t}$
production by the ATLAS Collaboration excluded a \Zp boson with a relative width of 3\%
with mass below 4.2~\TeV~\cite{EXOT-2018-49}, while the CMS Collaboration excluded a
\Zp boson with a relative width of 1\% with mass below 3.8~\TeV~\cite{CMS-B2G-17-017}.
Allowing for generic SM-like couplings of the \Zp boson, searches for
\Zp resonances decaying to dilepton final states exclude masses below 5.5~\TeV~\cite{EXOT-2025-09,CMS-EXO-19-019}.

The interaction Lagrangian of the \Zp boson with the top-quark is given by:
\begin{equation}
\begin{split}
\mathcal{L}
&= \bar{t}\,\gamma_\mu \left( c_L P_L + c_R P_R \right) t\, Z'^{\mu} \\
&= c_t\, \bar{t}\,\gamma_\mu \left( \cos\theta\, P_L + \sin\theta\, P_R \right) t\, Z'^{\mu},
\end{split}
\end{equation}
where $\gamma_\mu$ are the Dirac matrices and $P_{L/R} = (1 \mp \gamma_5)/2$ are the chirality projection operators.
The chirality angle is defined as $\tan \theta = c_R/c_L$, where $c_L$ and $c_R$ denote the couplings to the left- and right-handed top quarks, respectively, and the coupling of $\Zp$ boson to top quark is defined as $c_t = \sqrt{c_L^2 + c_R^2}$.

The dominant contribution to $Z'$ production arises from $t\bar{t}Z'$ production. At tree level, its rate depends on $c_t^2$, since the left- and right-handed contributions do not interfere. Consequently, the cross-section is independent of $\theta$.
In contrast, \tZp production proceeds through electroweak single-top-like diagrams,
where the contribution is proportional to $\cos^2\theta$, vanishing at
$\theta = \pi/2$ and reaching its maximum at $\theta = 0$, where it
becomes comparable in magnitude to the \ttZp contribution.
Both production modes are included in this paper and the full $\theta$ range is explored.

This search targets final states containing either exactly two leptons with the same electric charge
or at least three leptons.\footnote{In this paper, leptons refer to either electrons or muons, including those coming from $\tau$-lepton decays.}
These signatures provide strong sensitivity due to the small level of SM background contamination.
The dominant backgrounds arise from \tttt production and from $t\bar{t}$ production in association with a $W$ ($t\bar{t}W$), $Z$ ($t\bar{t}Z$), or Higgs boson ($t\bar{t}H$).
The analysis uses $pp$ collision data recorded by the ATLAS detector at $\sqrt{s} =$13~\TeV between 2015 and 2018 and at $\sqrt{s}=$13.6~\TeV between 2022 and 2023.
The two data samples are analysed separately and then combined using a profile-likelihood fit.
The event processing is performed within \textsc{FastFrames}
framework~\cite{FastFrames}, and the statistical interpretation is
carried out using \textsc{TRExFitter}
framework~\cite{TRExFitter}. Limits on the $Z'$ production cross-section are presented for two
representative $\theta$ values ($0$ and $\pi/2$), corresponding to the
left-handed and right-handed coupling scenarios. Limits are also
presented as $c_t^2,\theta$ contour for two representative $Z'$ masses.

This paper is structured as follows. Section~\ref{sec:detector} describes the ATLAS detector. The data and simulated samples are presented in Section~\ref{sec:samples} while the object and event selections are detailed in Section~\ref{sec:eventSelection}. The analysis strategy is outlined in Section~\ref{sec:strategy} with the background estimation and signal-background discrimination described in Sections~\ref{sec:bkgEstimation} and ~\ref{sec:signalDiscrimination}, respectively.
Systematic uncertainties are discussed in Section~\ref{sec:systematic}, the results are presented in Section~\ref{sec:result}, and conclusions are given in Section~\ref{sec:conclusion}.

% End of text imported from the .//sections/1-introduction.tex input file

%
\section{ATLAS detector}
\label{sec:detector}

% The next lines are included from the .//sections/2-detector.tex input file

The ATLAS detector~\cite{PERF-2007-01,GENR-2019-02} at the LHC covers nearly the entire solid angle around the collision point.\footnote{\AtlasCoordFootnote}
It consists of an inner tracking detector surrounded by a thin superconducting solenoid, electromagnetic and hadron calorimeters,
and a muon spectrometer incorporating air-core toroidal magnets. The inner detector (ID) is immersed in a \SI{2}{\tesla} axial magnetic field
and provides charged-particle tracking in the range of \(|\eta| < 2.5\). It consists of silicon pixel, silicon microstrip, and transition radiation tracking detectors.
Lead/liquid-argon (LAr) sampling calorimeters provide electromagnetic (EM) energy measurements with high granularity within the region \(|\eta|< 3.2\). A steel/scintillator-tile hadronic calorimeter covers the
central pseudorapidity range (\(|\eta|< 1.7\)). The endcap and forward regions are instrumented with LAr calorimeters for EM and hadronic energy measurements up to \(|\eta|=4.9\). The muon spectrometer surrounds
the calorimeters and is based on three large superconducting air-core toroidal magnets with eight coils each. The field integral of the toroids ranges between 2.0 and 6.0~Tm across most of the detector. The muon spectrometer includes a system of precision tracking chambers up to \(|\eta|= 2.7\) and fast detectors for triggering up to \(|\eta|= 2.4\).
The luminosity is measured primarily using the LUCID–2 detector, which is located
close to the beampipe.
Interesting events are selected by the first-level trigger system implemented in custom hardware,
followed by selections made by algorithms implemented in software in the high-level trigger~\cite{TRIG-2016-01,TRIG-2022-01}.
The first-level trigger accepts events from the \SI{40}{\MHz} bunch crossings at a rate below \SI{100}{\kHz},
which the high-level trigger further reduces to \SI{1.25}{\kHz} and \SI{3}{\kHz} on average in Run~2 and Run~3, respectively.

An extensive software suite~\cite{ATL-SOFT-PUB-2021-001} is used for the simulation, reconstruction and analysis of real
and simulated data, in detector operations, and in the trigger and data acquisition systems of the experiment.

% End of text imported from the .//sections/2-detector.tex input file

%
\section{Data and simulated event samples}
\label{sec:samples}

% The next lines are included from the .//sections/3-samples.tex input file
The data were collected using single-lepton triggers in both Run~2 and Run~3, complemented in Run~3 by dilepton triggers,
with requirements on the identification, isolation, and transverse momentum ($p_\textrm{T}$) across the full momentum range to
control the trigger rates~\cite{TRIG-2016-01,TRIG-2018-05,TRIG-2018-01,TRIG-2022-01}.
For single-lepton triggers, $p_\textrm{T}$ thresholds of 26, 60, and 140~\GeV were used for electrons, whereas for muons the thresholds were $p_\textrm{T} = 26$ and
50~\GeV.\footnote{For 2015 data, the electron trigger thresholds were 24, 60, and 120~\GeV. For 2022 and 2023 data, a lower threshold of 24~\GeV was used for muons.}
For dilepton triggers, the electron $p_\textrm{T}$ thresholds were 17 or 24~\GeV, while the muon $p_\textrm{T}$ threshold was 14~\GeV.
The data are required to satisfy the data quality criteria described in Ref.~\cite{DAPR-2018-01}.
The corresponding integrated luminosity is $140$~\ifb\ for Run~2 ~\cite{DAPR-2021-01} and $56$~\ifb\ for Run~3.

Signal and background processes were modelled using Monte Carlo (MC) samples.
These were processed through a simulation of the ATLAS detector geometry and response based on \GEANT~\cite{Agostinelli:2002hh,SOFT-2010-01}.
Corrections are applied to simulated events to ensure that object reconstruction and identification efficiencies,
energy scales, and resolutions, match those measured in data.
The generator versions quoted below apply to both Run~2 and Run~3; when they differ, the Run~3 value is given in parentheses after the Run~2 value.

The \Zp signal processes were generated at leading-order (LO) in QCD using \MG 3.5.1~\cite{Alwall:2011uj}
in the five-flavour scheme with the \NNPDF[3.1nlo] parton distribution function (PDF) set~\cite{Ball:2014uwa}.
The generated events were interfaced with \pythiaeight.309~\cite{Sjostrand:2014zea,Bierlich:2022pfr} for parton shower (PS),
hadronisation using the A14 tune.
The model implementation follows the simplified top-philic resonance framework described in Ref.~\cite{Kim_2016}.
The renormalisation ($\mu_R$) and factorisation ($\mu_F$) scales are set to $\mu_R=\mu_F=m_\textrm{T}/2$, where $m_\textrm{T}$ is the scalar sum of the transverse masses $\sqrt{m^2+\pTsq}$ of the particles generated from the matrix element (ME) calculation.
Signal samples were generated for resonance masses between 400~\GeV and 3000~\GeV.
All top quarks were required to be on shell at the matrix-element level.
The nominal configuration assumes a coupling strength $c_t = 3$ and a chirality parameter $\theta = \pi/4$, corresponding to equal left-handed and right-handed couplings.
Alternative coupling values ($c_t = 1, 4$) and chirality angles ($\theta = 0, \pi/2$) are considered and obtained via \MG ME reweighting~\cite{Mattelaer:2016gcx}.
The relative width of the \Zp resonance varies from approximately 4\% for $c_t = 1$\ at 400~\GeV to 35\% for $c_t = 3$ at 3000~\GeV.
For all signal samples, both $s$-channel and subdominant $t$-channel contributions are included.
Within the analysis selection, the interference between $\ttZp$ and the SM \tttt\ process is evaluated and found to have negligible impact on both the cross-section and kinematic distributions.

The SM \tttt\ process was simulated with \MGNLO[3.5.3]~\cite{Frederix:2012ps,Alwall:2014hca} generator, which provides MEs at next-to-leading-order~(NLO) in the strong coupling constant \alphas with the \NNPDF[3.1nlo] PDF set.
The $\mu_R$ and $\mu_F$ scales are set to $\mu_R=\mu_F=m_\textrm{T}/4$, following Ref.~\cite{Frederix:2017wme}.
Top-quark decays were simulated at LO using \MADSPIN~\cite{Frixione:2007zp,Artoisenet:2012st} to preserve spin correlations.
Events were interfaced with \pythiaeight.230 (8.310) for PS and hadronisation using the A14 tune.
An alternative \tttt\ sample was generated with \HERWIG[7.0.4] (7.2.3)~\cite{Bahr:2008pv,Bellm:2015jjp,Bellm:2019zci} with the H7UE tune~\cite{Bellm:2015jjp} to assess the uncertainties associated with the PS and hadronisation model.
A further sample generated at NLO with the \Sherpa[2.2.11] (2.2.14)~\cite{Bothmann:2019yzt} was used to evaluate the uncertainty related to the choice of event generator.
For this sample, the scales are set to $\mu_R=\mu_F=H_\textrm{T}/2$, where $H_\textrm{T}$ was defined as the scalar sum of the transverse momenta of all final-state particles.
The \tttt\ samples are normalised to a cross-section of 10.15~fb and 12.0~fb for Run~2 and Run~3, respectively, computed at NLL$'$ accuracy and matched to the complete NLO prediction in QCD including electroweak (EW) corrections~\cite{Melissa:4top}.

The \ttW process was modelled using the \Sherpa[2.2.10] (2.2.14) generator with the \NNPDF[3.0nnlo] PDF set.
The MEs were calculated for up to one additional parton at NLO and up to two partons at LO\@.
The additional partons were matched and merged with the \sherpa PS using Catani--Seymour dipole factorisation~\cite{Schumann:2007mg} and the MEPS@NLO prescription~\cite{Hoeche:2011fd,Hoeche:2012yf,Catani:2001cc,Hoeche:2009rj}, with a merging scale of 30~\GeV.
Virtual QCD corrections at NLO accuracy were provided by the \OPENLOOPS[2]~\cite{Cascioli:2011va,Denner:2016kdg,Buccioni:2019sur} library.
The renormalisation and factorisation scales are set to $\mu_R=\mu_F=m_\textrm{T}/2$.
Event-by-event correction factors are applied to account for virtual NLO EW corrections at $\mathcal{O}(\alpha^{2}\alphas^{2})$ and LO corrections at $\mathcal{O}(\alpha^{3})$~\cite{Bothmann:2019yzt,Kallweit:2015dum,Gutschow:2018tuk}.
The \ttW~samples are normalised to a cross-section of 745.3~fb, computed at NLO in QCD for Run~2.
The ratio of the theory prediction to the generator cross-section obtained for Run~2 is applied to the Run~3 samples.
In addition, sub-leading EW corrections at $\mathcal{O}(\alpha^{3}\alphas)$~\cite{Frederix:2017wme} are included via an additional \Sherpa[2.2.10] sample produced at LO for Run~2 and via event-by-event correction factors for Run~3.
An alternative \ttW~sample generated with \MGNLO, including up to one additional parton at NLO, is used to evaluate the generator uncertainty.
In this sample, jet multiplicities were merged using the FxFx NLO ME and PS merging prescription~\cite{Frederix:2012ps}.
A dedicated \MGNLO \ttW sample including the sub-leading EW corrections to $\mathcal{O}(\alpha^{3}\alphas)$ is also considered.

Simulated $t\bar{t}Z$ events were generated using \MGNLO at NLO, including off-shell $Z$ contributions with $m(\ell^+\ell^-) >$ 1~\GeV (5~\GeV) for Run~2 (Run~3).
The samples are normalised to a prediction at NLO QCD+EW~\cite{deFlorian:2016spz}.
Alternative samples were produced with \MGNLO interfaced to \HERWIG~7, and with the A14 Var3c up and down tune variations.
The $\ttbar H$ process was modelled using the \POWHEGBOX~\cite{Frixione:2007nw,Nason:2004rx,Frixione:2007vw,Alioli:2010xd,Hartanto:2015uka} generator, which provides MEs at NLO in the strong coupling constant \alphas in the five-flavour scheme with the \NNPDF[3.0nlo] PDF set.
The cross-section is normalised to the calculation at NLO in QCD and EW accuracy using \MGNLO, as reported in Ref.~\cite{deFlorian:2016spz}.
The \ttbar and single-top-quark samples were modelled using the \POWHEGBOX[v2]~\cite{Frixione:2007nw,Nason:2004rx,Frixione:2007vw,Alioli:2010xd} generator at NLO in QCD with the \NNPDF[3.0nlo] PDF set.
For \ttbar production, the \hdamp parameter, which controls the transverse momentum \pt of the first additional emission beyond the Born configuration, is set to 1.5\,\mtop~\cite{ATL-PHYS-PUB-2016-020}.
The overlap between \ttbar and $tW$ productions is removed using the diagram removal scheme~\cite{Frixione:2008yi}.
The $\ttt$ sample was generated at LO in QCD using the five-flavour scheme and includes $\tttW$ and $\tttj$ processes separately.
These samples are normalised to NLO cross-sections computed with \MGNLO,
resulting in 1.606 (1.936)~fb for $\tttW$ with a scale choice of $\HT/8$,
and in 0.73 (0.839)~fb for $\tttj$ with a scale choice of $\HT/4$ for Run~2 (Run~3).
Rare background contributions were modelled using \MGNLO for $tWZ$, $tZ$, $t\bar{t}VV$, $\ttHH$ and $\ttWH$, \POWHEGBOX for $VH$, and \Sherpa for $V$+jets, $VV$ and $VVV$.
These are then normalised to their respective NLO theoretical cross-sections.

% End of text imported from the .//sections/3-samples.tex input file

%
\section{Event reconstruction and selection}
\label{sec:eventSelection}

% The next lines are included from the .//sections/4-selection.tex input file
Events are required to contain at least one primary vertex (PV) reconstructed from at least two ID tracks, each with $\pT>0.5~\GeV$.
If more than one PV candidate satisfies these criteria, then the PV with the largest sum of $\pT^2$~ over all associated ID tracks is selected~\cite{ATL-PHYS-PUB-2015-026,IDTR-2021-01}.

Electron candidates are reconstructed from energy deposits in the EM calorimeter matched to tracks in the ID~\cite{EGAM-2018-01}
and are required to satisfy the `TightLH' identification criterion~\cite{PERF-2017-01}.
Selected electrons must have transverse momentum $\pT>15~\GeV$ and pseudorapidity $|\eta|<2.47$, excluding the transition region between the barrel and the endcap calorimeters ($1.37<|\eta| < 1.52$).
Electrons are required to be well isolated using the \texttt{Tight\_VarRad} working point, based on criteria applied to topological clusters in the calorimeter and ID tracks~\cite{EGAM-2021-01}.

Muon candidates are reconstructed by combining tracks reconstructed in the ID with tracks in the muon spectrometer~\cite{PERF-2015-10,MUON-2018-03}.
They are required to satisfy $\pT>15~\GeV$ and $|\eta|<2.5$. The `Medium' muon identification working point is used.
Muons are further required to satisfy isolation requirements based on the properties of ID tracks around the reconstructed muon, using the \texttt{PflowTight} working point~\cite{MUON-2018-03}.

Electron (muon) candidates are matched to the primary vertex by requiring that the significance of the transverse impact parameter, $d_0$, satisfies $|d_0|/\sigma(d_0)<5 (3)$, where $\sigma(d_0)$ is the measured uncertainty in $d_0$, and that the longitudinal impact parameter, $z_0$, satisfies $|z_0\sin{\theta}|<0.5$~mm.
Electrons and muons satisfying all criteria described above are referred to as ``tight''.
Leptons satisfying the same criteria without the isolation requirements are referred to as “loose”.

Jets are reconstructed from particle-flow~\cite{PERF-2015-09} objects using the anti-$k_t$ algorithm~\cite{Cacciari:2008gp,Fastjet} with the radius parameter $R=0.4$.
Jets are calibrated using simulation with in situ corrections derived from data~\cite{JETM-2018-05},
and are required to have $\pT> 20$~\GeV\ and be within $|\eta|<2.5$.
To reduce the effect of pile-up, the jet vertex tagger (JVT)~\cite{PERF-2014-03} is applied to jets with $\pT<60~\GeV$ and $|\eta|<2.4$.

Jets containing $b$-hadrons are identified ($b$-tagged) using the GN2 tagger~\cite{FTAG-2023-05},
a transformer-based neural network designed to classify the flavour of reconstructed jets.
A jet is considered $b$-tagged if it passes the operating point (OP) corresponding to an average efficiency of 85\% for jets containing $b$-hadrons in simulated $t\bar{t}$ events.
Four additional OPs are available with average $b$-tagging efficiencies of 90\%, 77\%, 70\%, and 65\%.
To fully exploit the $b$-tagging information, a pseudo-continuous $b$-tagging (PCBT) score is assigned to each jet,
indicating the tightest operating point satisfied. Jets are assigned an integer score from 2 to 6 accordingly, with jets failing all OPs assigned a score of 1.

A sequential overlap removal procedure based on loose leptons is applied to ensure that the same calorimeter energy deposit or track is not used in more than one reconstructed object, as described in Ref.~\cite{TOPQ-2018-05}.
The missing transverse momentum, with magnitude \Etmiss, is defined as the negative vector sum of the transverse momenta of all reconstructed objects in the event.
In addition, a soft term is included to account for contributions from tracks originating from the primary vertex but not associated with any reconstructed object~\cite{JETM-2020-03}.

Events are required to have either a same-sign (SS) tight lepton pair (SS2L) or at least three tight leptons (3L).
At least one selected lepton must be matched to a trigger object, and the leading lepton must have $\pT > 28~\GeV$ to ensure high trigger efficiency.
In the SS2L channel, an electron charge identification (ECID) selection is applied in Run~2 to suppress background from electron charge misidentification.
This selection is based on a boosted decision tree (BDT) discriminant, rejecting approximately 90\% of electrons with misidentified charge while retaining 98\% of electrons with correctly measured charge~\cite{EGAM-2018-01}.
No ECID selection is applied for Run~3, as the selection is only available for Run~2.
Events with two same-sign electrons are required to have the invariant mass satisfying $m_{ee}>15~\GeV$ and $|m_{ee}-91~\GeV|>10~\GeV$, to suppress contributions from $Z\to ee$ events in which one electron has its charge misidentified.
In events with at least three leptons, all opposite-sign, same-flavour leptons pairs are required to satisfy $|m_{\ell\ell}-91~\GeV|>10~\GeV$ to reduce the contributions from $Z$ boson decays.

% End of text imported from the .//sections/4-selection.tex input file

%
\section{Search strategy}
\label{sec:strategy}

% The next lines are included from the .//sections/5-strategy.tex input file
A signal-enriched region is defined by exploiting the high jet and $b$-jet multiplicities, and the large overall transverse momentum of jets and leptons in \ttZp and \tZp events.
Events are required to contain at least six jets, of which at least two are $b$-tagged, and to satisfy $\HT > 500$~\GeV, where $\HT$ is defined as the scalar sum of the transverse momenta of the selected leptons and jets.
This region is referred to as the inclusive signal region (SR).

To further enhance the sensitivity, the inclusive SR is subdivided into multiple categories based on the $b$-jet and lepton multiplicities, as summarised in Table~\ref{tab:subregions}.
A mass-parameterised transformer-based classifier is trained on events in the inclusive SR to
discriminate the signal from the SM backgrounds, as detailed in Section~\ref{sec:signalDiscrimination}.

Dedicated control regions (CRs) are defined to constrain the dominant \ttW and non-prompt lepton backgrounds.
A simultaneous profile likelihood fit is performed across all regions to extract the \Zp signal.
An overview of the SR and CRs, together with the fitted observables used in each region, is provided in Table~\ref{tab:regions}.

\begin{table}[t]
\centering
\caption{ Definition of analysis SR categories based on the number of $b$-tagged jets $(N_{b})$ and the number of leptons $(N_{\ell})$.}
\label{tab:subregions}
\begin{tabular}{l l l}
\toprule \hline \noalign{\smallskip}
Region         & $N_{b}$   & $N_{\ell}$    \\
\noalign{\smallskip} \hline
\noalign{\smallskip}
SR 2$b$2$\ell$       & $=2$    & $=2$      \\
SR 2$b$3$\ell$       & $=2$    & $\geq 3$ \\
SR 3$b$2$\ell$       & $=3$    & $=2$     \\
SR 3$b$3$\ell$       & $=3$    & $\geq3$  \\
SR 4$b$              & $\geq4$ & $\geq2$  \\
\noalign{\smallskip} \hline \noalign{\smallskip} \bottomrule
\end{tabular}

\end{table}

\begin{table}[!htbp]
\centering
\caption{Definition of the signal and control regions. The first column shows the region name as used in the text. The event selection requirements are defined in the middle columns. The last column shows the fitted variable in each region. The ``Channel'' column specifies the lepton multiplicity and flavour composition. $N_j$ ($N_b$) indicates the jet ($b$-tagged jet) multiplicity, $H_\textrm{T}$ is defined as the scalar sum of the transverse momenta of the selected leptons and jets. $\ell_1$, $\ell_2$, and $\ell_3$ refer to the highest $p_\textrm{T}$, second-highest $p_\textrm{T}$, and third-highest $p_\textrm{T}$ leptons, respectively. $D_{\mathrm{signal}}$ denotes the transformer discriminant trained to separate signal from the background.}
\small
\begin{tabular}{c c c c l c}
\toprule \hline \noalign{\smallskip}
Region      & Channel       & $N_{j}$     & $N_{b}$     & Other selection  & \begin{tabular}[c]{@{}l@{}}Fitted\\variable\end{tabular}\\
\noalign{\smallskip} \hline
\noalign{\smallskip}
SR      & SS2L,3L    & $\geq 6$      & $\geq 2$      & $H_\textrm{T} > 500$~\GeV & $D_{\mathrm{signal}}$ \\ \hline
\multirow{5}{*}{CR $\ttW^{+/-}$} & \multirow{5}{*}{SS$e\mu$, $\mu\mu$} & \multirow{5}{*}{$\geq 4$} & \multirow{5}{*}{$\geq 2$}     & \rule{0pt}{2.6ex}$\pT^{\ell_2}> 25$~\GeV, $|\eta(e)| < 1.37$ & \\
&   &   &   & $N_b = 2$: $\HT < 500$~\GeV or $N_j < 6$ & \\
&   &   &   & $N_b \geq 3$: $\HT < 500$~\GeV & $N_j$  \\
&   &   &   & CR $\ttW^+$: total lepton charge $> 0$ & \\
&   &   &   & CR $\ttW^-$: total lepton charge $< 0$ & \\ \hline
\multirow{4}{*}{CR 1b(+/-)} &  \multirow{4}{*}{SS2L, 3L} & \multirow{4}{*}{$\geq 4$} & \multirow{4}{*}{$ = 1$} &  \rule{0pt}{2.6ex}$\pT^{\ell_2} > 25$~\GeV   & \\
&   &   &   & $\HT > 500$~\GeV                             &  $N_j$ \\
&   &   &   & CR 1b(+): total lepton charge $> 0$                   & \\
&   &   &   & CR 1b(-): total lepton charge $< 0$                   & \\ \hline
\multirow{2}{*}{CR Low $m_{\gamma^*}$}      & \multirow{2}{*}{SS$ee$, $e\mu$} & \multirow{2}{*}{$4 \leq N_j < 6$} & \multirow{2}{*}{$\geq 1$} & Only one $e$ from $\gamma^\ast \rightarrow ee$  & Yield\\
&   &   &   & No $e$ from $\gamma \rightarrow ee$ \\ \hline
\multirow{2}{*}{CR Mat. Conv.}              & \multirow{2}{*}{SS$ee$, $e\mu$} & \multirow{2}{*}{$4 \leq N_j < 6$} & \multirow{2}{*}{$\geq 1$} & Only one $e$ from $\gamma \rightarrow ee$       & Yield \\
&   &   &   & No $e$ from $\gamma^\ast \rightarrow ee$ \\ \hline
\multirow{3}{*}{CR HF$\mu$} & \multirow{3}{*}{$e\mu\mu$, $\mu\mu\mu$} & \multirow{3}{*}{$\geq 1$}     & \multirow{3}{*}{$= 1$}    & $100 < \HT < 300$~\GeV & \\
&   &   &   & $\MET > 50$~\GeV  & $\pT^{\ell_3}$ \\
&   &   &   & total lepton charge $= \pm 1$ & \\ \hline
\multirow{3}{*}{CR HF$e$} & \multirow{3}{*}{$eee$, $ee\mu$} & \multirow{3}{*}{$\geq 1$}  & \multirow{3}{*}{$= 1$}    & $100 < \HT < 275$~\GeV & \\
&   &   &   & $\MET > 35$~\GeV & $\pT^{\ell_3}$\\
&   &   &   & total lepton charge $= \pm 1$ & \\
\noalign{\smallskip} \hline \noalign{\smallskip} \bottomrule
\end{tabular}
\label{tab:regions}
\end{table}

% End of text imported from the .//sections/5-strategy.tex input file

\subsection{Background estimation}
\label{sec:bkgEstimation}

% The next lines are included from the .//sections/5-1-background.tex input file
Background processes are categorised according to the origin of the selected leptons as
either prompt or non-prompt. Prompt leptons typically originate from
$W$ or $Z$ boson decays, with dominant contributions
from \tttt, \ttW, \ttZ, and $t\bar{t}H$ production. Smaller contributions include $VV$, $VVV$,
and $VH$ processes, and rare processes such as $t\bar{t}VV$, $t\bar{t}t$, $tZq$, and $tWZ$.
The prompt background is estimated by using MC simulation as described
in Section~\ref{sec:samples}. The predicted background normalisations are used
except for \ttW, which is constrained using data in dedicated CRs.

The non-prompt background comprises events containing at least one fake/non-prompt lepton, originating
from heavy-flavour hadron decays, photon conversions, or jets
misidentified as leptons. The QmisID background comprises events
containing an electron with misidentified lepton charge. These backgrounds originate primarily
from $t\bar{t}$, $V$+jets, and single-top production. The fake/non-prompt component is estimated
together with \ttW using a template method, while the QmisID component is estimated by using
data-driven techniques. The QmisID contribution affects only electrons, as the charge
misidentification rate for muons is negligible.

The template method uses simulation to model the kinematic
distributions of the fake/non-prompt lepton and \ttW backgrounds,
while their normalisations are constrained in dedicated CRs.
These CRs are included in the likelihood fit together with the SRs, and the normalisation factors are
determined simultaneously with the \Zp signal normalisation.

Four CRs are defined to determine the \ttW normalisation: CR $t\bar{t}W^+$, CR $t\bar{t}W^-$, CR 1b(+), and CR 1b(-), as summarised in Table~\ref{tab:regions}.
The ``$+$'' (``$-$'') regions correspond to events with positive (negative) total charge, defined as the sum of the electric charges of the selected leptons.
The CRs $t\bar{t}W^\pm$ are defined using $e\mu$ and $\mu\mu$ events with at least four jets, of which at least two are $b$-tagged.
Additional requirements on \HT are applied to ensure orthogonality with the SRs.
The $ee$ channel is excluded due to its large contamination from QmisID\@.
The remaining QmisID contribution in the $e\mu$ channel is reduced by requiring electrons to satisfy $|\eta| < $ 1.37.
The CRs 1b($\pm$) include both SS2L and 3L events with the same jet requirements but require exactly one $b$-tagged jet.
A requirement of $\HT >$ 500~\GeV is imposed to better match the kinematic properties of the SRs.
Pre-fit distributions of number of jets, $N_j$, used in the template method are shown in Figure~\ref{fig:prefit-ttWCR} for all four CRs.

\begin{figure}[t]
\begin{center}
\includegraphics[width=0.9\linewidth]{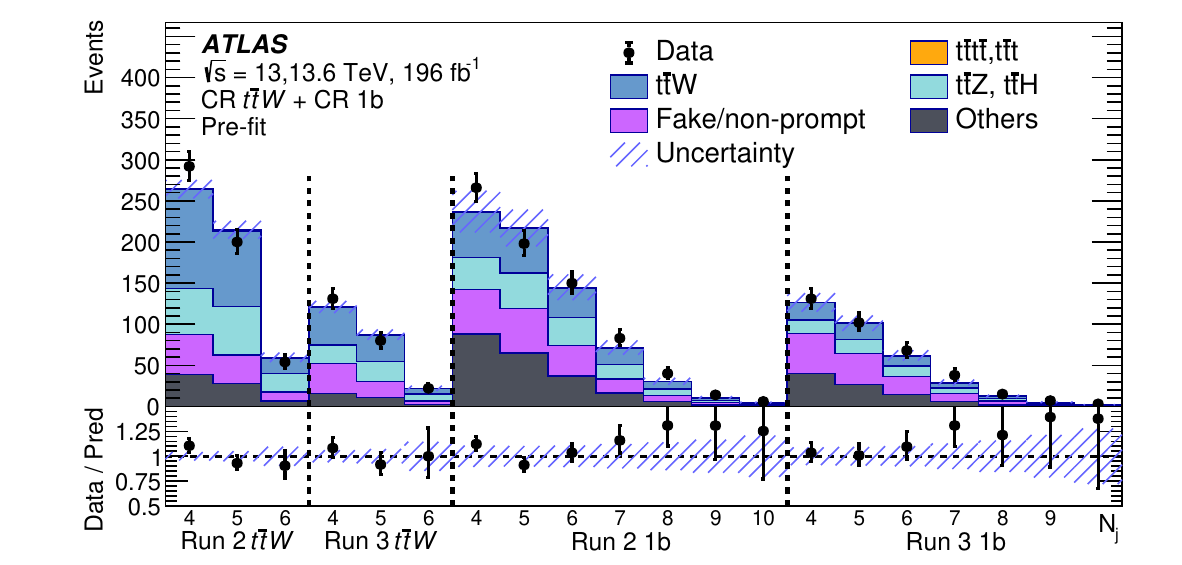} \hspace{0.5cm}
\caption{\label{fig:prefit-ttWCR}Pre-fit distributions of the jet multiplicity ($N_j$) in
CRs $t\bar{t}W^\pm$ and 1b($\pm$), shown separately for Run~2 and Run~3.
Events are shown in each CR, as indicated by the labels along the $x$-axis.
The stacked histograms represent the SM background predictions, while the markers show the observed data.
The lower panel shows the ratio of data to the total pre-fit prediction, with the
hatched band indicating the total uncertainty. }
\end{center}
\end{figure}

The non-prompt background is categorised into four main components based on the true particle origin of the reconstructed leptons:
\begin{itemize}
\item Low $m_{\gamma^*}$: events with a virtual photon producing an $e^+e^-$ pair;
\item Mat. Conv.: events with one non-prompt electron originating from photon conversion in the detector material;
\item HF$e$: events with one non-prompt electron from heavy-flavour decay;
\item HF$\mu$: events with one non-prompt muon from heavy-flavour decay;
\end{itemize}
Dedicated CRs are defined to constrain the normalisation of each component: CR Low $m_{\gamma^*}$,
CR Mat. Conv., CR HF$e$, and CR HF$\mu$, as summarised in Table~\ref{tab:regions}.

The Low $m_{\gamma^*}$ and Mat. Conv. CRs are defined using $ee$ and $e\mu$ events with $4 \leq N_j < 6$, of which at least one is $b$-tagged.
Electrons from photon conversions are identified through their small invariant mass with a nearby reconstructed track.
They are classified as an internal conversion, $\gamma^\ast \rightarrow ee$, if the associated track points to the
primary vertex, and as material conversion, $\gamma \rightarrow ee$, if the track originates from
a conversion vertex in the detector material.
CR Low $m_{\gamma^*}$ requires one electron from internal conversion,
while CR Mat. Conv. selects events with one material conversion electron.
For Run~3, the CRs Low $m_{\gamma^*}$ and Mat. Conv. are not used.
Without ECID selection in Run~3, these regions become dominated by QmisID and
are therefore no longer suitable to constrain the corresponding background components.
In the Run~3-only fit, the normalisation factors for these components are obtained from the Run~2 results,
as the underlying processes and detector condition are similar between Run~2 and Run~3.
In the combined Run~2 and Run~3 fit, the normalisation factors are primarily constrained by the Run~2 CRs,
with correlations applied between the two data samples.

The CR HF$e$ (HF$\mu$) is defined by selecting events with three leptons,
namely $eee$ and $ee\mu$ ($e\mu\mu$ and $\mu\mu\mu$), and exactly one $b$-tagged jet.
The selections target $t\bar{t}$ dileptonic decays with an extra non-prompt lepton in events
with low \HT.
Figure~\ref{fig:prefit-fakeCRs} shows the pre-fit distributions used in the template method, including the third-highest lepton $p_\textrm{T}$ and the event yields.

\begin{figure}[t]
\begin{center}
\subfloat[]{\includegraphics[width=0.66\linewidth]{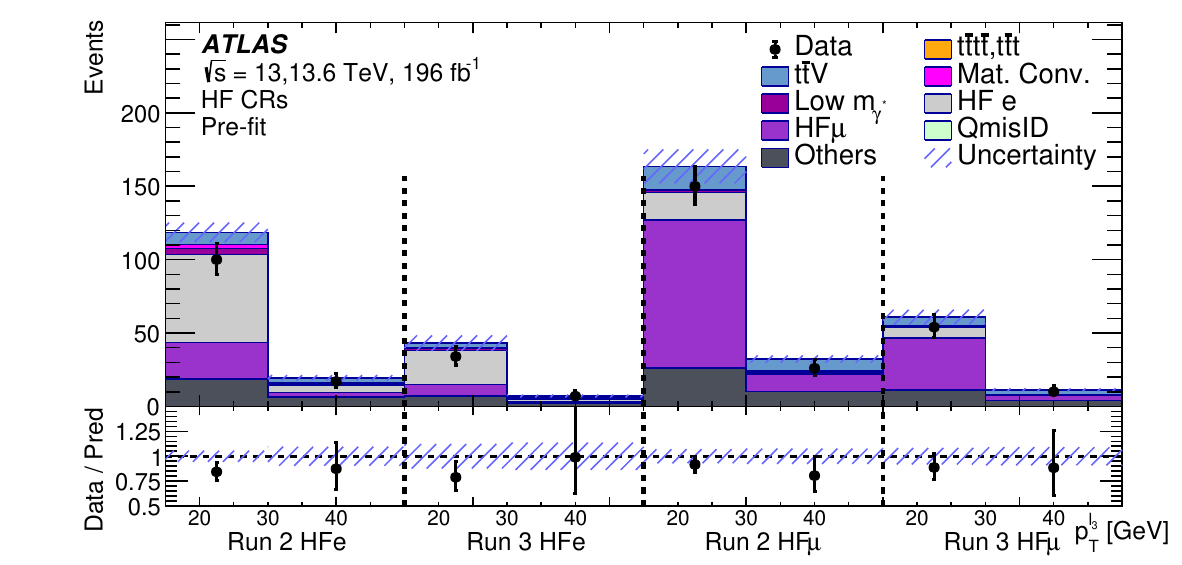}}
\subfloat[]{\includegraphics[width=0.355\linewidth]{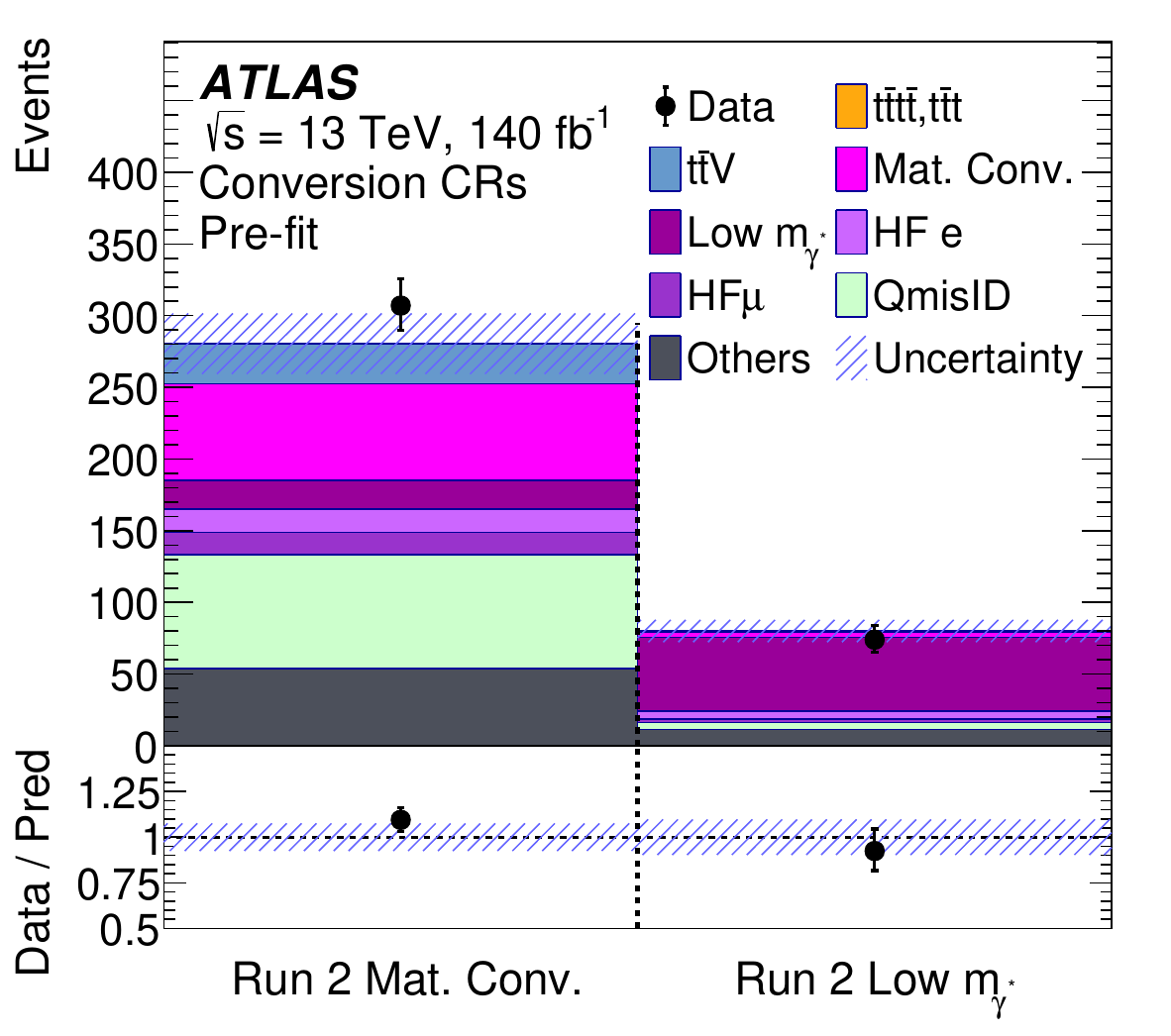}}
\caption{\label{fig:prefit-fakeCRs}Pre-fit distributions of the fitted variables in the CRs targeting non-prompt background.
In (a), CR HF$e$ and CR HF$\mu$, the third-highest lepton $p_\textrm{T}$ is used as the fitted variable for Run~2 and Run~3 data.
In (b), CR Mat. Conv. and CR Low $m_{\gamma^*}$, the event yields are shown for Run~2 data; these regions are not defined for Run~3.
The stacked histograms represent the SM background predictions, while the markers show the observed data.
The lower panels show the ratio of data to the total pre-fit prediction, with the hatched band indicating the total uncertainty. }
\end{center}
\end{figure}

The QmisID background contributes only to the same-sign $ee$ and $e\mu$ channels.
It arises from misreconstruction of the electron charge,
either due to bremsstrahlung photon emission followed by conversion or from  mismeasurement of the track curvature.
The charge misidentification rate is measured using data samples of $Z \rightarrow ee$ events,
selected by requiring the invariant mass of the electron pair to be within 10~\GeV of the $Z$
boson mass, without imposing any charge requirement~\cite{EGAM-2018-01}.
For Run~3, the rate is parameterised as a function of the electron \pt and $\eta$.
In Run~2, the misidentification rate is reduced due to the ECID
selection; however, a residual dependence on the jet multiplicity
is observed and taken into account.
The rate varies from 0.006\% (0.02\%) to 4\% (10\%) for Run~2 (Run~3).
The expected QmisID yield is obtained by applying the measured misidentification rate to opposite-sign dilepton data events.

% End of text imported from the .//sections/5-1-background.tex input file

\subsection{Signal-to-background discrimination}
\label{sec:signalDiscrimination}

% The next lines are included from the .//sections/5-2-sigDiscrimination.tex input file
The background in the SR is dominated by \tttt and top-quark pair production in association with a boson.
To enhance the separation between signal and background events, a transformer-based neural network implemented in
the \textsc{Salt} framework~\cite{salt2025} is trained on events in the inclusive SR\@.
The model employs an attention mechanism~\cite{FTAG-2023-05,vaswani2023at}
and processes both event-level (global) and object-level features from jets and leptons.
The event-level inputs include $N_{j}$, $N_{b}$,
$N_{\text{leptons}}$, $H_{\text{T}}$, the missing transverse momentum $\MET$,
and its azimuthal angle $\ensuremath{\phi_{E_{\text{T}}^{\text{miss}}}}$.
The object-level inputs consist of \pT, $\eta$, $\phi$, and particle type,
supplemented by the charge for leptons and the PCBT score for jets.
The resonance mass hypothesis is included as an input feature and parameterised during training:
for signal events, the true mass is used, while for background events
it is randomly sampled from the set of simulated signal mass
hypotheses. Representative kinematic distributions in the inclusive SR
are shown in Figure~\ref{fig:incSR}.

The combined features are passed through an initialisation network,
whose output serves as input to a four-layer transformer encoder. Each encoder layer
comprises a multi-head attention block followed by a feed-forward network~\cite{vaswani2023at}.
The model is trained using a cross-entropy loss function without applying MC event weights.
Overtraining is mitigated using dropout and $k$-fold cross-validation with $k=4$, in which
the training sample is randomly divided into four subsets. In each iteration, two subsets are used
for training, one for validation, and one for testing.
The training targets correspond to \ttZp, \tZp, multi-top (\tttt and \ttt), $t\bar{t}V$ (\ttW, \ttH, and \ttZ), and other backgrounds.
A single discriminant is constructed from the softmax output probabilities:
\begin{equation}
D_{\mathrm{signal}} =
\log\left(
\frac{p_{\ttZp} + p_{\tZp}}{p_{\mathrm{multi-top}} + p_{t\bar{t}V} + p_{\mathrm{other\, bkg}}}
\right),
\end{equation}
where $p_i$ denotes the predicted probability for class $i$.

\begin{figure}[t]
\begin{center}
\subfloat[]{\includegraphics[width=0.45\linewidth]{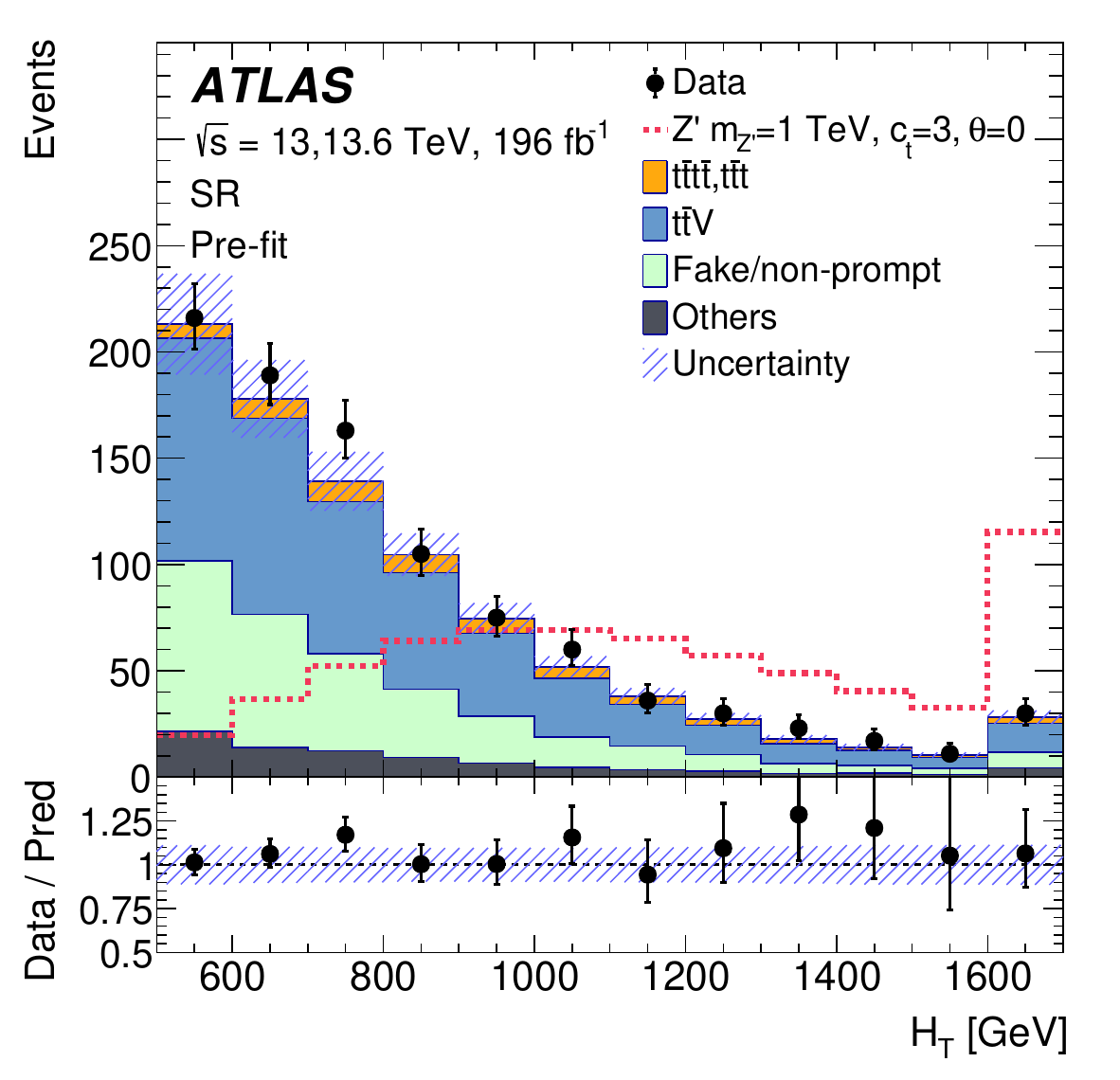}} \hspace{0.5cm}
\subfloat[]{\includegraphics[width=0.45\linewidth]{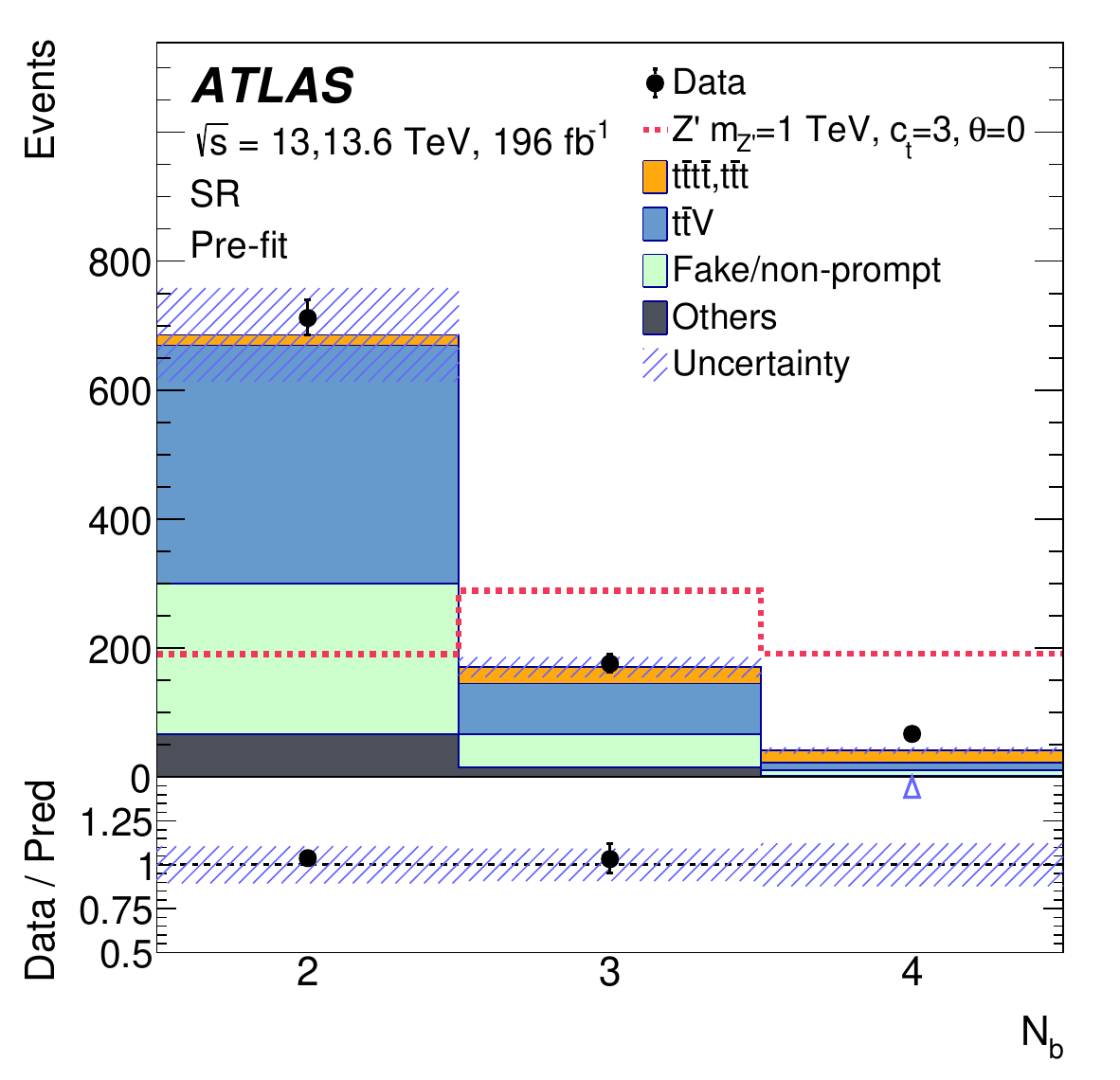}} \\
\subfloat[]{\includegraphics[width=0.45\linewidth]{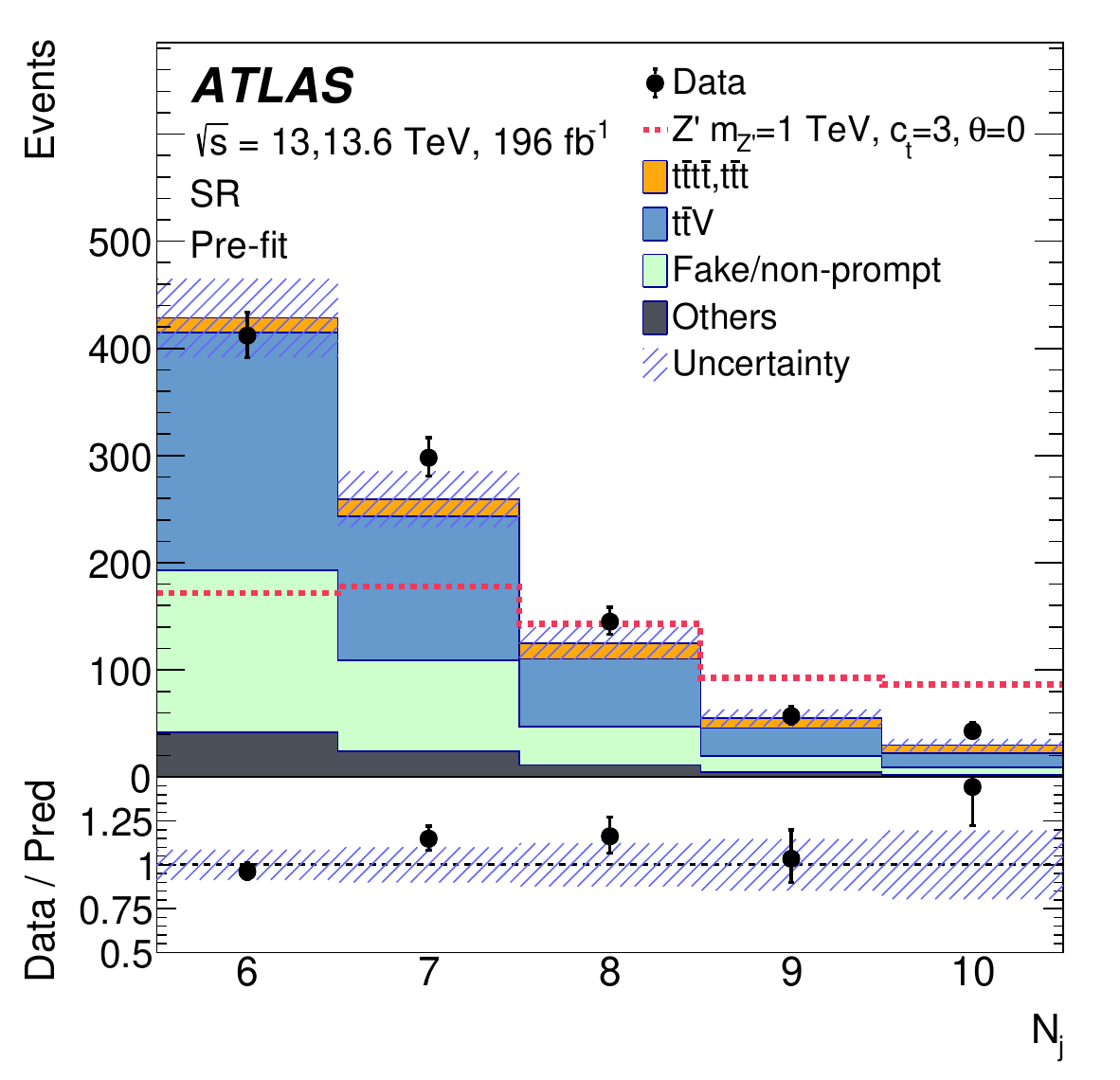}} \hspace{0.5cm}
\subfloat[]{\includegraphics[width=0.45\linewidth]{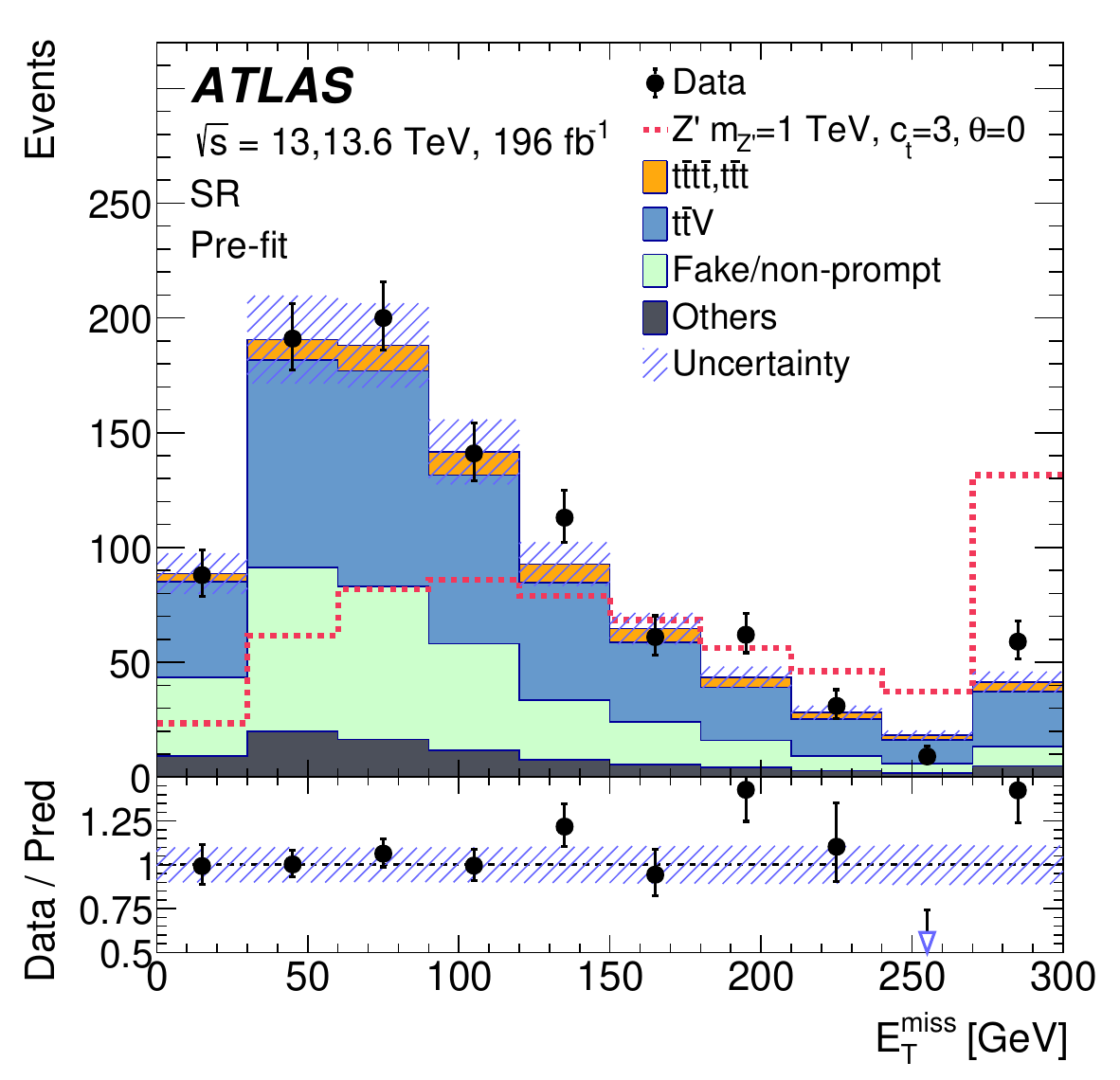}} \\
\caption{\label{fig:incSR} Pre-fit distributions in the inclusive SR, combining Run~2 and Run~3 data:
(a) \HT, (b) number of $b$-tagged jets, (c) jet multiplicity, and (d) $\MET$.
The stacked histograms represent the pre-fit SM background predictions and the markers indicate the observed data.
The lower panels show the ratio of data to the pre-fit predictions.
The expectation for a 1000~\GeV \Zp signal with $c_t = 3$ and $\theta = 0$ is overlaid using the predicted signal cross-section to illustrate the shape difference between the signal and the backgrounds.
In all distributions, the last bin includes the overflow events. }
\end{center}
\end{figure}

% End of text imported from the .//sections/5-2-sigDiscrimination.tex input file

%
\section{Systematic uncertainties}
\label{sec:systematic}

% The next lines are included from the .//sections/6-systematics.tex input file
Various sources of systematic uncertainty affect the estimated signal and background rates,
and the shape of the fitted variables. These can be classified into
experimental uncertainties, modelling uncertainties of physics processes, and
uncertainties arising from the estimate of reducible background.
In the combined fit to Run~2 and Run~3 data, most experimental uncertainties are
treated as uncorrelated, while all other systematic uncertainties are assumed to be correlated, unless stated otherwise.

\subsection{Experimental uncertainties}
The dominant experimental uncertainties arise from the jet energy scale (JES) and jet energy resolution (JER).
The JES and its associated uncertainty are derived using information from MC simulations, data taken during LHC runs, and test-beam data~\cite{JETM-2018-05}.\
The JES uncertainty consists of multiple correlated components, and a simplified set-up with an eigenvector decomposition is used.
A similar approach is used for the JER uncertainty. The Run~3 jet calibration is derived from the same
in situ measurement as in Run~2; therefore the uncertainties are treated as fully correlated between Run~2 and Run~3,
except for components arising from non-closure effects.
An uncertainty in the JVT efficiency is also included~\cite{PERF-2014-03}.

Uncertainties arising from the measurement of the $b$-tagging efficiencies and mis-tagging rates for $c$- and light-flavour jets~\cite{FTAG-2018-01,FTAG-2020-08,FTAG-2019-02,FTAG-2023-10,FTAG-2024-07} are subdominant.
Their impact is evaluated separately for each jet flavour in bins of jet \pT and $\eta$, and across all PCBT score bins.

Other experimental uncertainties have a smaller impact.
The uncertainty in the integrated luminosity amounts to 0.83\% for the Run~2 data~\cite{DAPR-2010-01} and 2\% for the Run~3 data~\cite{ATL-DAPR-PUB-2024-001}.
An uncertainty in the correction applied to the pile-up profile in the simulated samples is also considered.
Electron and muon uncertainties arise from the calibration of simulation to data, including corrections to trigger, reconstruction, identification, and isolation efficiencies, as well as to energy scale and resolution~\cite{MUON-2018-03,EGAM-2021-01,MUON-2022-01,EGAM-2021-02}.
An additional uncertainty associated with the ECID efficiency is considered for electrons in Run~2.
Finally, an uncertainty in \MET from possible miscalibration of its soft-track component is assigned~\cite{JETM-2020-03}.

\subsection{Modelling uncertainties}
Uncertainties in the modelling of the signal arise from missing higher-order QCD corrections.
These are estimated by varying the renormalisation and factorisation scales by factors of two and one half
relative to their nominal values, corresponding to the maximal upward and downward variations.
This procedure is applied to the signal and to the main background processes, \tttt, \ttW, \ttH, and \ttZ~\cite{TOPQ-2021-08-witherratum}.

Uncertainties in the modelling of \tttt production have the dominant impact on the search.
Contributions from matrix-element and parton-shower modelling are evaluated by comparing the nominal prediction with alternative samples simulated using \SHERPA or \MGNLO interfaced to \HERWIG~7, respectively.
A theoretical uncertainty derived from renormalisation and factorisation scale variations of the NLL$'$ prediction matched to the complete NLO calculation including QCD and electroweak corrections is assigned to the inclusive \tttt\ cross-section,
amounting to $^{+29\%}_{-24\%}$ at $\sqrt{s}=13$~\TeV and $^{+30\%}_{-24\%}$ at $\sqrt{s}=13.6$~\TeV~\cite{Melissa:4top}.
An additional shape uncertainty is assigned to account for EW and NLL' contributions, parameterised as a function of the \tttt invariant mass, following Ref.~\cite{Frederix:2017wme,Melissa:4top}.
The PDF uncertainty of 1\% is assigned, following Ref.~\cite{TOPQ-2021-08-witherratum} and based on the PDF4LHC prescription~\cite{Butterworth:2015oua}.

For \ttZ, uncertainties associated with the choice of parton shower are determined by comparing
the nominal MC prediction with a \MGNLO interfaced to \HERWIG~7, while uncertainties in the modelling of
QCD radiation are evaluated using the A14 Var3c up and down tune variations.
For \ttH, the nominal simulation is compared with samples simulated using \PYTHIA~8 with alternative \textsc{pThard} setting and \POWHEG interfaced to \HERWIG 7, respectively.
For \ttW, uncertainties associated with the modelling of additional QCD radiation, and the choice of
the ME generator and parton shower, are evaluated by comparing the nominal prediction with an
alternative sample simulated at NLO using \MGNLO for Run~2 and extrapolated to Run~3.
For all three processes, a PDF uncertainty of 1\% is assigned following the same prescription as in \tttt measurement~\cite{TOPQ-2021-08-witherratum}.

The cross-section uncertainties in backgrounds other than \tttt have a minor impact on the search.
Cross-section uncertainties of 12\% and 10\% are applied to the \ttZ and \ttH backgrounds~\cite{deFlorian:2016spz}.
No cross-section uncertainty is assigned to \ttW, as its normalisation is determined in the template fit.
An uncertainty of 30\% is applied to the sum of the \tZ and \tWZ processes~\cite{TOPQ-2012-20}. The uncertainty
in the \ttt production cross-section is set to 30\%~\cite{Frederix:2017wme}.
The uncertainty in $VV$ production is taken as the difference between the measured differential cross-section
as a function of the jet multiplicity and the prediction from \SHERPA 2.2.2~\cite{STDM-2018-03}.
For $VV$ events with $\leq 3$, $4$, and $\geq 5$ jets, uncertainties of 20\%, 50\%, and 60\% are applied, respectively, with the corresponding components treated as uncorrelated.
For all other minor background processes with negligible impact, a conservative uncertainty of 50\% is considered.

For all simulated background contributions, except \tttt, additional uncertainties are
considered for events containing additional $b$-jets. This is motivated by the expected
high $b$-jet multiplicities in signal-like events and the difficulty in modelling the production of extra $b$-jets not originating from the hard-scattering process.
For each background process, an uncorrelated uncertainty of 50\% is assigned to events with three
true $b$-jets\footnote{`true' $b$-jets are identified in simulated events as jets that are ghost-associated~\cite{Cacciari:2008} with $b$-hadrons.} and to events with at least four true $b$-jets.
These uncertainties are based on the measurement of $t\bar{t}$+jets production with additional heavy-flavour jets~\cite{TOPQ-2017-12} and comparisons between data and prediction in $t\bar{t}\gamma$ events with three and four $b$-tagged jets.
A variation of 100\% was tested and found to have a negligible impact on the results.

\subsection{Uncertainties in reducible backgrounds}
The uncertainties in QmisID background arise from the measurement of the electron charge misidentification rates.
The following contributions are considered: the statistical uncertainty from the likelihood fit to data;
the variation in the measured rate evaluated by varying dielectron invariant mass requirement;
and the non-closure uncertainty taken as the relative difference between the observed and the predicted misidentification rate.
The resulting uncertainties range from 4.3\% (3.5\%) to 34\% (28\%) for Run~2 (Run~3), depending on the lepton $p_\textrm{T}$ and $\eta$, and additionally on $N_j$ in Run~2.

The uncertainties in the heavy-flavour non-prompt backgrounds arise solely from the shape of the distributions used as fit discriminants in every analysis region.
These uncertainties are estimated from a bin-by-bin comparison of data and the post-fit background prediction with the loose lepton selection.
The uncertainties in the photon-conversion backgrounds are estimated in a region enriched in $Z(\to \mu\mu) + \gamma$ events by comparing data and simulation.
In addition, a conservative uncertainty of 100\% is assigned to leptons from light-flavour sources~\cite{TOPQ-2018-05}.

% End of text imported from the .//sections/6-systematics.tex input file

%
\section{Results}
\label{sec:result}

% The next lines are included from the .//sections/7-results.tex input file
The search results are obtained from a profile-likelihood fit~\cite{Rolke:2004mj} performed simultaneously
across all signal and control regions at both centre-of-mass energies. The fit is carried out for each signal
mass and coupling hypothesis separately, and for the background-only hypothesis.
Systematic uncertainties affecting signal and background processes are incorporated
into the likelihood function as Gaussian nuisance parameters
correlated across all bins and regions, while MC statistical
uncertainties are incorporated through Poisson constraints in each
bin. The parameter of interest is the signal strength $\mu$, a
common normalisation factor to
the signal yield in all bins~\cite{Cowan:2010js-err}.
In the absence any significant excess, upper exclusion limits are set at 95\% confidence level (CL) on the cross-section
times branching ratio in Run~2 and Run~3, and on the signal strength
in the combined fit, thus accounting for the slight difference in
the predicted signal cross-section for the different centre-of-mass energies. The
limits are evaluated with the modified frequentist ${\mathrm{CL}}_{\mathrm{s}}$ method using the asymptotic approximation~\cite{Read:2002hq,Junk:1999kv}.
The binning of the fitted variable in the SRs, $D_{\mathrm{signal}}$, is optimised independently for
each signal hypothesis to maximise the separation between signal and background,
while avoiding bins with large statistical uncertainties in the dominant background predictions.

Under the background-only hypothesis, the normalisation factors for different background processes,
extracted from the fit to data, are shown in Table~\ref{tab:NFs}.
The fit results are obtained using the mass-parameterised discriminant evaluated at a signal mass of 1000~\GeV.
Studies performed at alternative signal mass hypotheses show negligible changes in the fitted background normalisation
factors, indicating that the results depend only weakly on the chosen mass point.
Good agreement between data and post-fit background is observed in all CRs, as illustrated in
Figure~\ref{fig:postfitCRs}. The goodness-of-fit evalauted using a saturated model yields a probability of 65.5\%.
The normalisation factor for \ttW, $\text{NF}_{\ttW}$,
is consistent with the dedicated \ttW measurement using Run~2 data~\cite{TOPQ-2016-11}.
The modelling of the $\ttW$ background is further validated in a dedicated
validation region, where the expected charge asymmetry is confirmed by comparing
event yields with positive and negative sums of lepton charges.

\begin{table}[htp!]
\begin{center}
\caption{ The normalisation factors for \ttW and non-prompt background processes determined from the background-only fit. The fit is done simultaneously in all signal and control regions. The uncertainties include both statistical and systematic uncertainties.}
\begin{tabular}{c|ccccc}
\toprule
Parameter &    $\text{NF}_{\ttW}$    &    $\text{NF}_{\text{Mat.\ Conv.}}$  & $\text{NF}_{\text{Low }m_{\gamma^*}}$ & $\text{NF}_{\text{HF}e}$ & $\text{NF}_{\text{HF}\mu }$ \\
\midrule
Run~2     & $1.09^{+0.15}_{-0.14}$   &    $1.40^{+0.40}_{-0.35}$            &  $0.87^{+0.22}_{-0.19}$               & $0.72^{+0.21}_{-0.20}$    & $0.89^{+0.17}_{-0.16}$  \\
\midrule
Run~3     & $1.20^{+0.34}_{-0.27}$   &    -       &      -           &  $0.64^{+0.33}_{-0.28}$ & $0.79^{+0.26}_{-0.23}$ \\
\midrule
Combined & $1.11^{+0.13}_{-0.13}$    &    $1.42^{+0.38}_{-0.34}$       &      $0.89^{+0.22}_{-0.19}$           &  $0.69^{+0.18}_{-0.17}$ & $0.89^{+0.14}_{-0.13}$ \\
\bottomrule
\end{tabular}
\label{tab:NFs}
\end{center}
\end{table}

\begin{figure}[t]
\begin{center}
\subfloat[]{\includegraphics[width=0.64\linewidth]{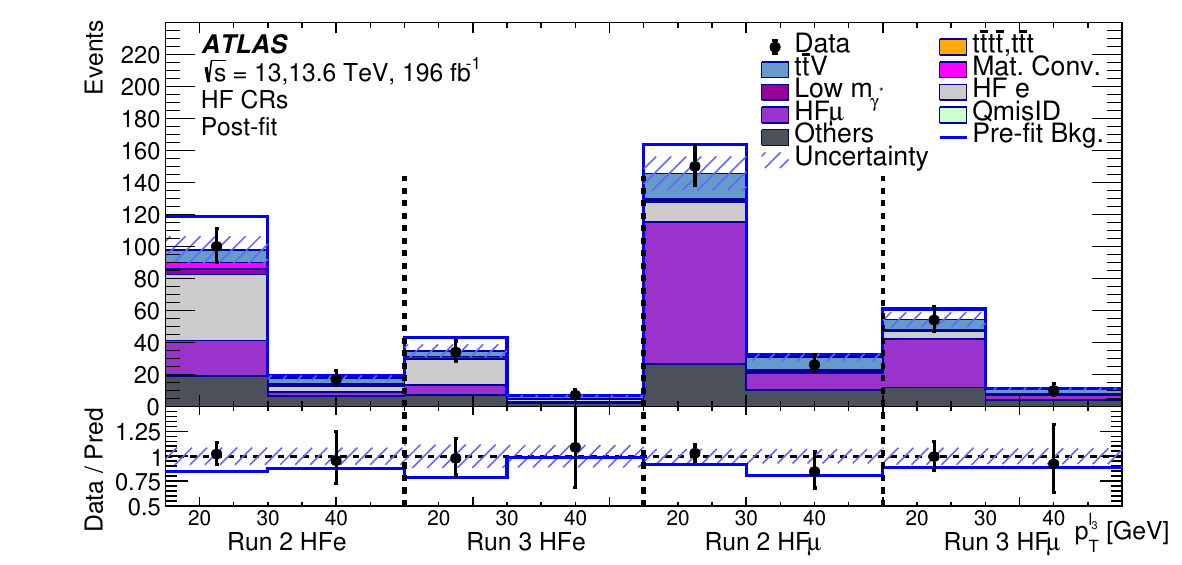}}
\subfloat[]{\includegraphics[width=0.345\linewidth]{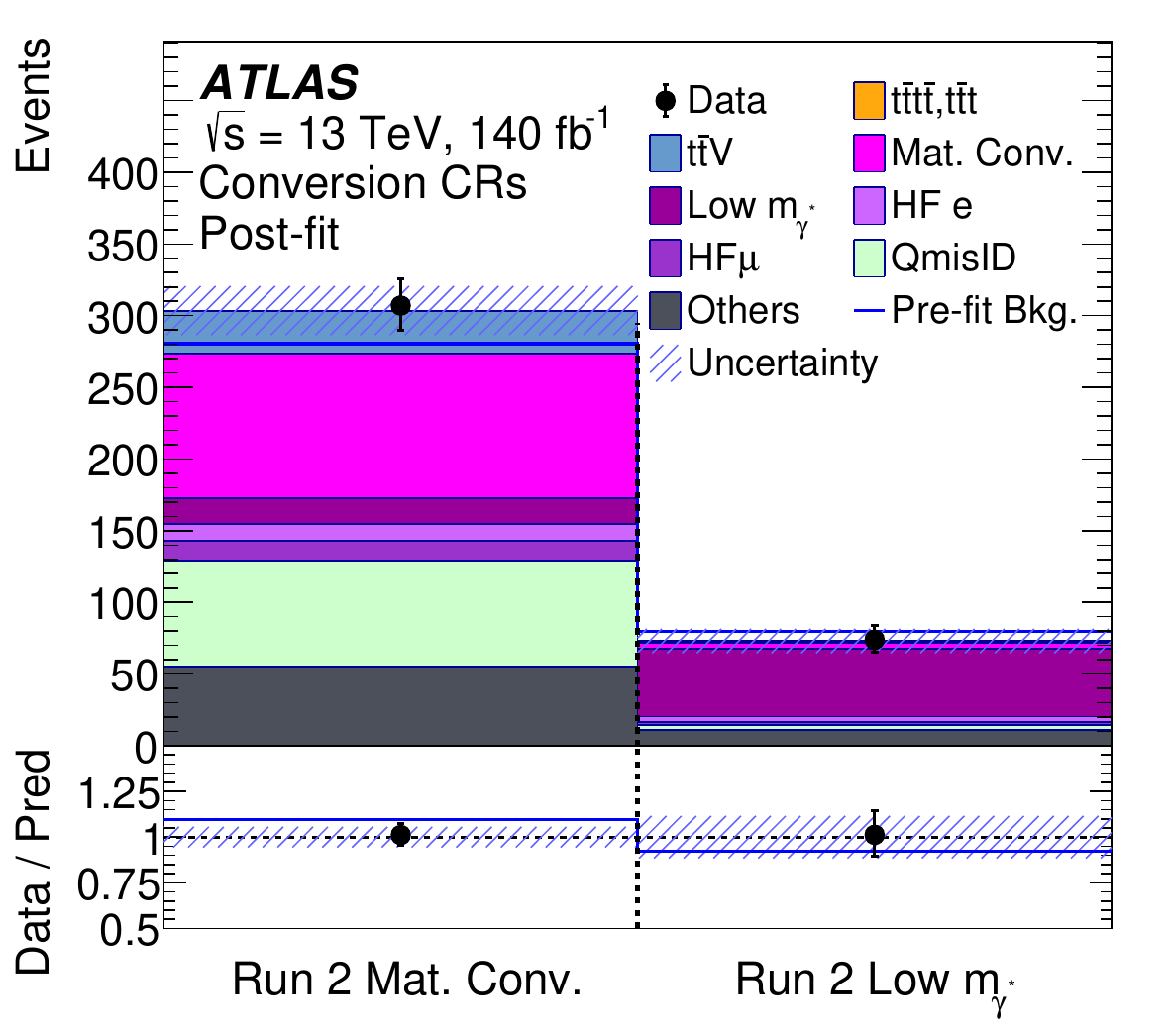}} \\
\subfloat[]{\includegraphics[width=0.9\linewidth]{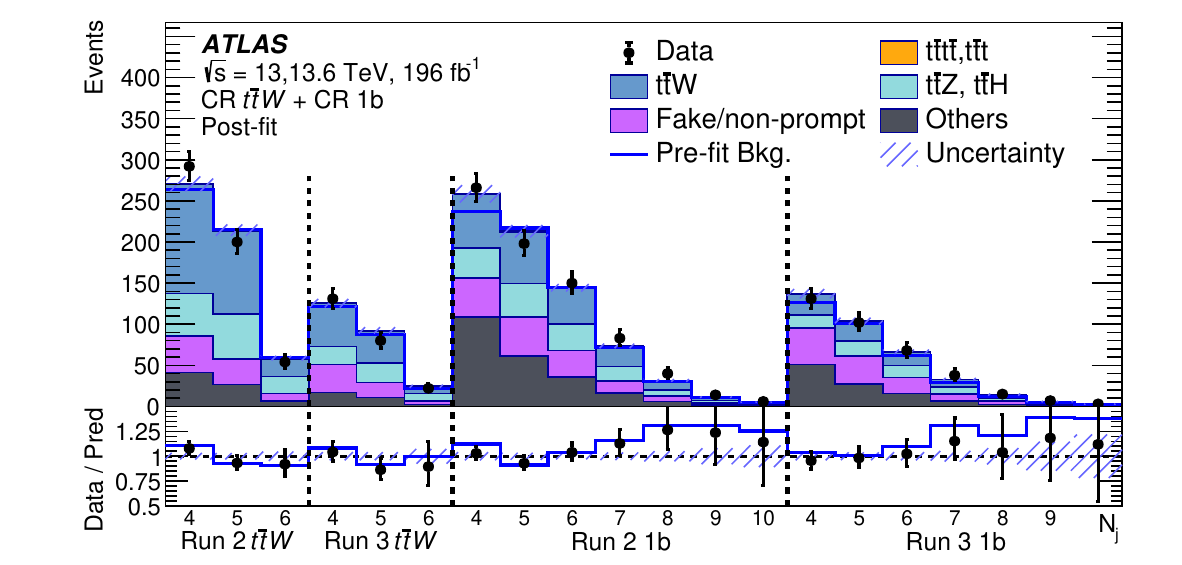}}
\caption{\label{fig:postfitCRs} Post-fit distributions of the fitted variables in the CRs, obtained under the background-only hypothesis: (a) HF$e$ and HF$\mu$, (b) photon conversion CRs, and (c) all ttW CRs. The stacked histograms represent the post-fit background predictions, with the hatched band indicating the total uncertainty. The pre-fit prediction is overlaid for comparison. The lower panels show the ratio of data to the predictions, where the dots correspond to the post-fit predictions and the line to the pre-fit prediction.}
\end{center}
\end{figure}

Figure~\ref{fig:postfitSRs} shows the post-fit distribution of
$D_{\mathrm{signal}}$ in the SRs for the background-only fit,
separately for Run~2 and Run~3.
As the multivariate discriminant is parameterised as a function of the \Zp mass,
the shape of $D_{\mathrm{signal}}$ varies across the signal hypotheses;
the distributions shown correspond to the 1000~\GeV mass point.
The binning shown is identical to that used in the corresponding
signal-plus-background fits and is optimised according to the expected
signal-to-background discrimination, with adjacent bins merged where their
separation power is limited. Overall, good agreement between
data and the fitted background prediction is observed across all regions.
The compatibility of the data with the background-only hypothesis is
quantified using a $p$-value, scanning twelve mass points, three
$\theta$ values and two $c_t$ values.
The largest excess over the background-only expectation is observed
at $m_{Z'}=400$~\GeV, $c_t = 3$, and $\theta = \pi/4$, with a $p$-value of 0.009, corresponding to a local significance of 2.4 $\sigma$.

\begin{figure}[t]
\begin{center}
\subfloat[]{\includegraphics[width=0.9\linewidth]{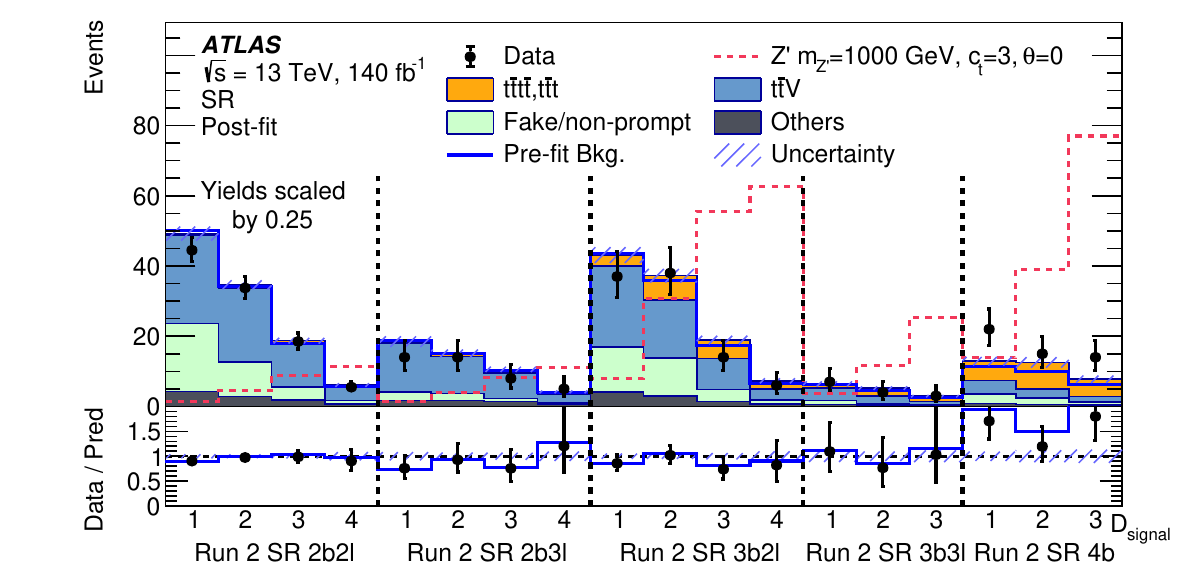}} \\
\subfloat[]{\includegraphics[width=0.9\linewidth]{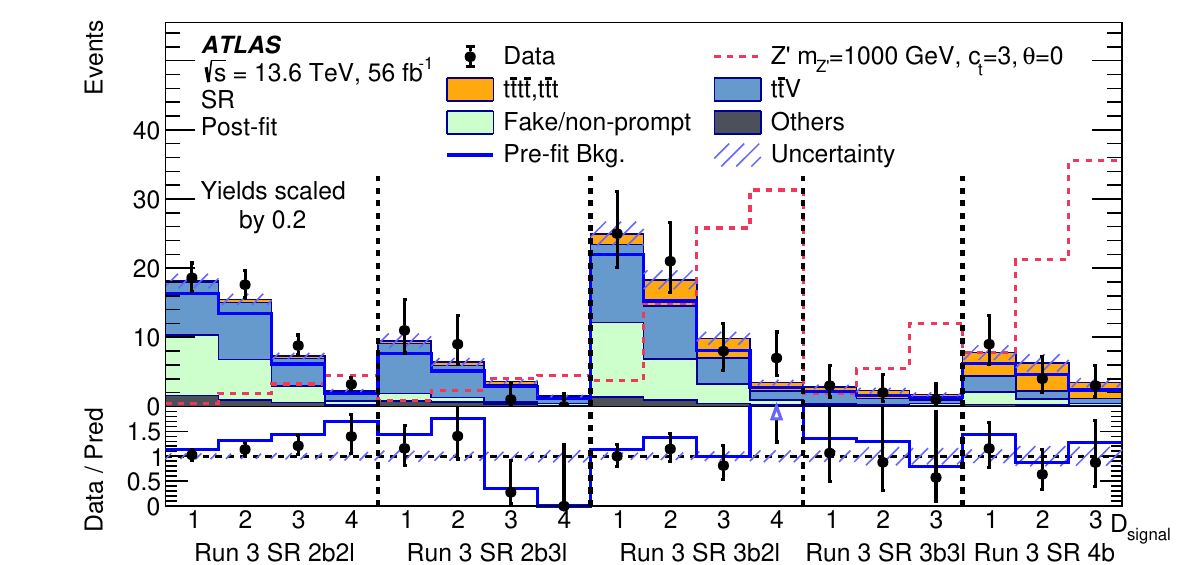}}
\caption{\label{fig:postfitSRs} Distribution of the $D_{\mathrm{signal}}$ in the SRs for data compared to the post-fit
background prediction under the background-only hypothesis, obtained from the combined Run 2 and Run 3 fit.
The data and corresponding predictions are shown separately for (a) Run~2 and (b) Run~3.
The contributions from the different background processes are indicated, together with the expected signal for $m_{\Zp} = $ 1000~\GeV with
$c_t = 3$ and $\theta = 0$. The vertical dashed lines separate the
different SRs, as labelled along the horizontal axis. Within each region, the
discriminant bins are labelled as discrete bin indices. The lower panels show the ratio of data to the prediction, where the dots correspond to the post-fit predictions and the line to the pre-fit prediction.
}
\end{center}
\end{figure}

Constraints on the signal are derived under the signal-plus-background hypothesis.
The resulting upper limits on the signal strength are presented as a function of
the \Zp boson mass for the $c_t = 1, 3$ and $\theta = 0, \pi/2$ scenarios in Figure~\ref{fig:limit1dCombined}.
Lower limits are set on the mass of the hypothetical \Zp~bosons by
comparing the observed limit on the signal strength to the theoretical
prediction, using a linear interpolation between MC sample points.
For $c_t = 3$, \Zp masses below 2000~\GeV (2200~\GeV expected) are excluded
at $95$\% CL for $\theta = 0$, while masses below
1750~\GeV (1900~\GeV expected) are excluded for $\theta = \pi/2$.
For $c_t = 1$, \Zp masses below 1150~\GeV  (1250~GeV expected) are excluded
at $95$\% CL for $\theta = 0$, while masses below
900~\GeV (1000~GeV expected) are excluded for $\theta = \pi/2$.
The addition of Run~3 data increases the sensitivity by 30\% at 1.5~\TeV, with larger gains at higher \Zp mass.
The observed cross-section limits present an improvement by about a factor 15 over the previous ATLAS \ttZp search~\cite{EXOT-2022-14},
which was not able to rule out coupling $c_t = 1$ for any \Zp mass
using events containing a single-lepton and boosted large-radius jets.
The results can also be compared to a CMS search for a top-philic \Zp boson in the single-lepton channel using the full Run~2 data~\cite{CMS-B2G-24-009}.
For the 4\% fixed-width benchmark, which matches the width of the
$c_t=1$ model considered here, the cross-section limit observed by CMS
is comparable at $m_{\Zp}= 3$~\TeV and about a factor 20 worse than
this analysis at $m_{\Zp}= 500$~\GeV.

A summary of the impact of the different sources of systematic uncertainty
on the signal strength, for representative values for the \Zp mass
1000~\GeV with $c_t = 1$ and $\theta = 0$, is provided in
Table~\ref{tab:groupedSyst}.
The decomposition of the total uncertainty into statistical and systematic components
is performed using a BLUE-like method based on the post-fit
covariance matrix assuming Gaussian-distributed systematic uncertainties~\cite{Pinto:2023yob}.
This search is dominated by statistical uncertainty.
Among the systematic uncertainties, the largest contributions arise from the  modelling of the \tttt and \ttW
processes. The fitted signal strength is partially anti-correlated
with the \tttt cross-section and generator choice uncertainties, with
a linear correlation coefficients of $-0.37$ and $-0.29$ at 400~\GeV, and
$-0.26$ and $-0.21$ at 1000~\GeV.
Among the experimental uncertainties, the dominant contributions arise from the jet and $b$-tagging calibrations.

To assess the impact of SM \tttt cross-section uncertainty, an alternative fit is performed
with an asymmetric upward variation of 70\%, motivated by the higher cross-section
observed in the ATLAS Run~2 SM \tttt measurement compared to the theoretical prediction~\cite{TOPQ-2021-08-witherratum}.
For a mass of 1000~\GeV, $c_t = 1$ and $\theta = 0$, the correlation of the signal strength with the
nuisance parameter for the \tttt cross-section uncertainty in this configuration is about 53\%,
compared to 26\% in the nominal fit. The nuisance parameter is constrained and shifted up to 90\%$\pm$60\%
of its initial value, compared to 50\%$\pm$95\% in the nominal fit. These competing effects of the
larger background estimate and larger background uncertainty result in unchanged observed limit on the signal.

\begin{figure}[t]
\begin{center}
\subfloat[]{\includegraphics[width=0.49\linewidth]{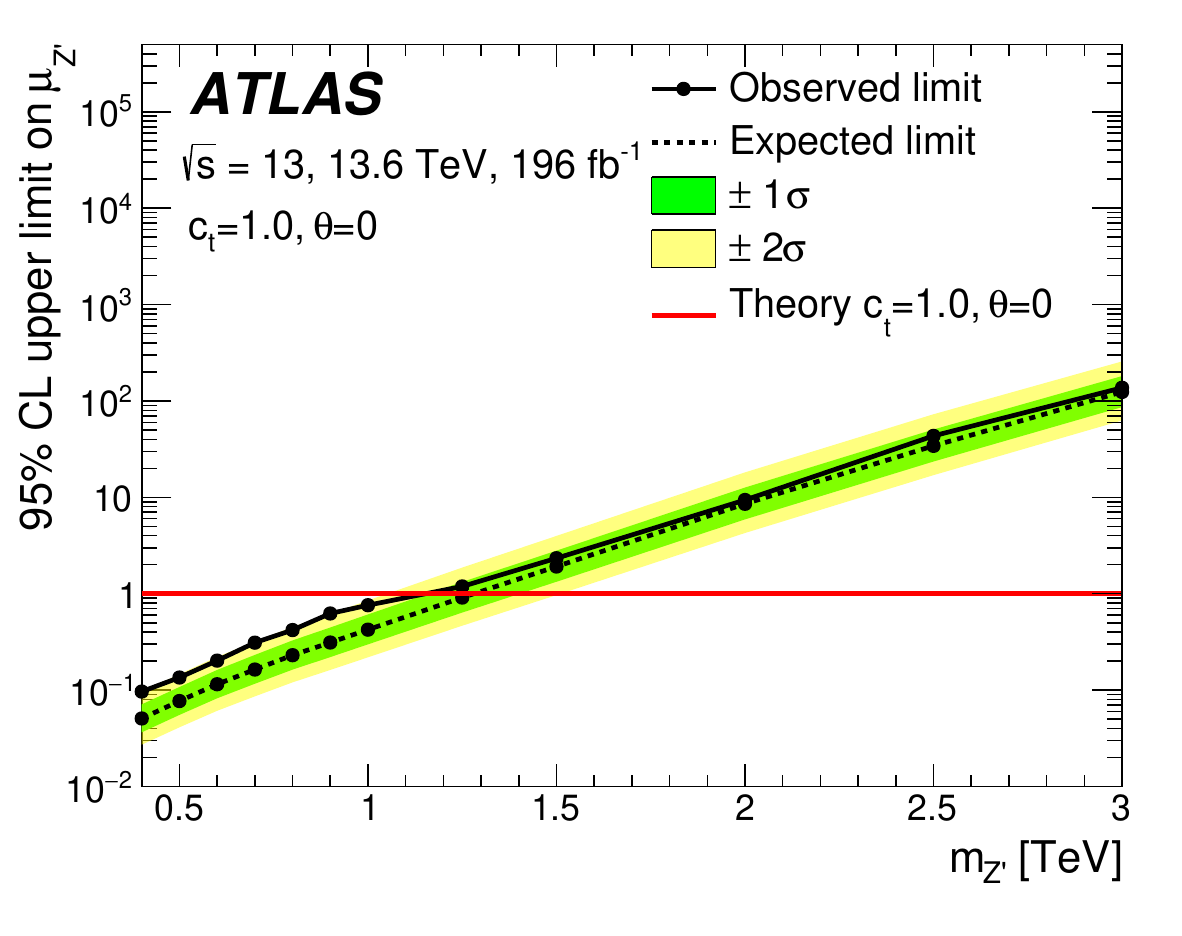}}
\subfloat[]{\includegraphics[width=0.49\linewidth]{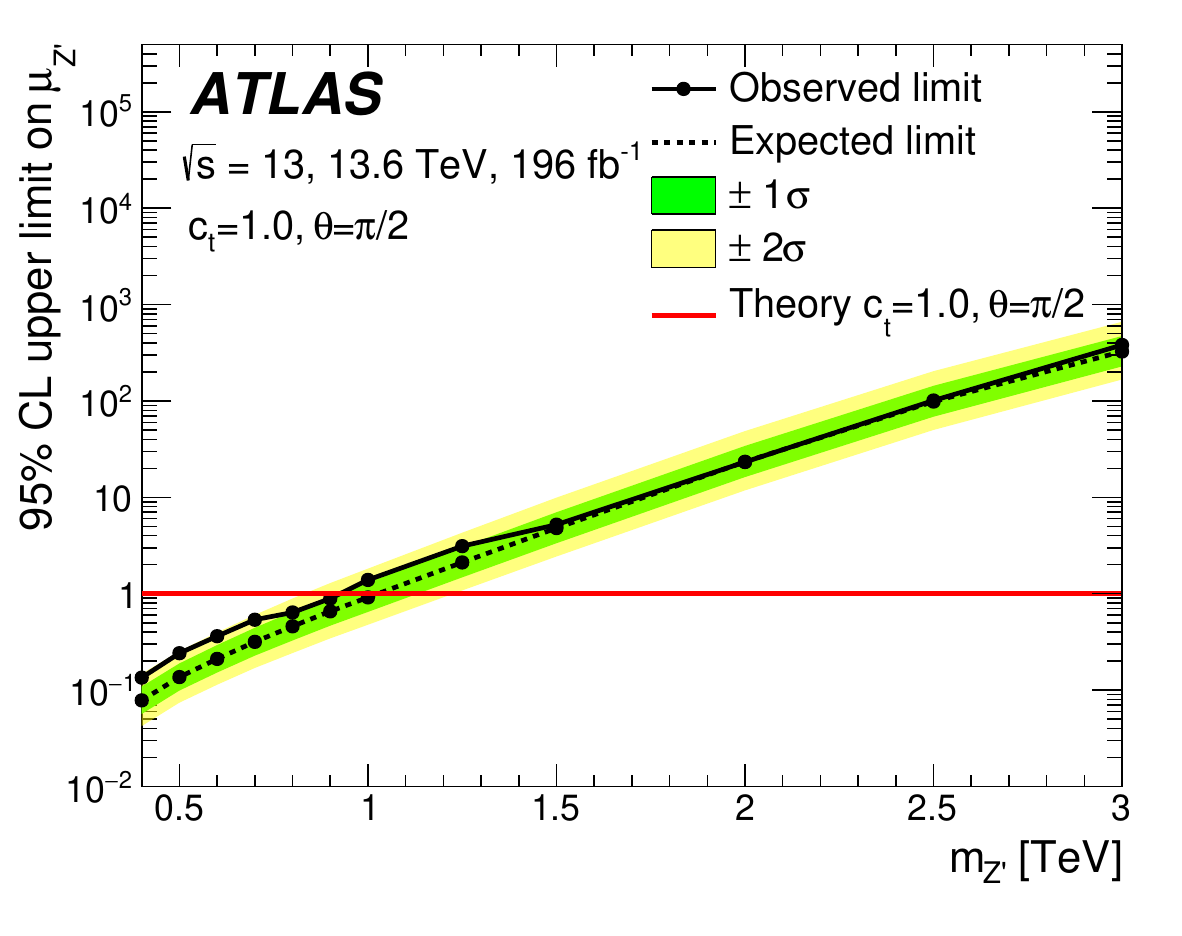}}\\
\subfloat[]{\includegraphics[width=0.49\linewidth]{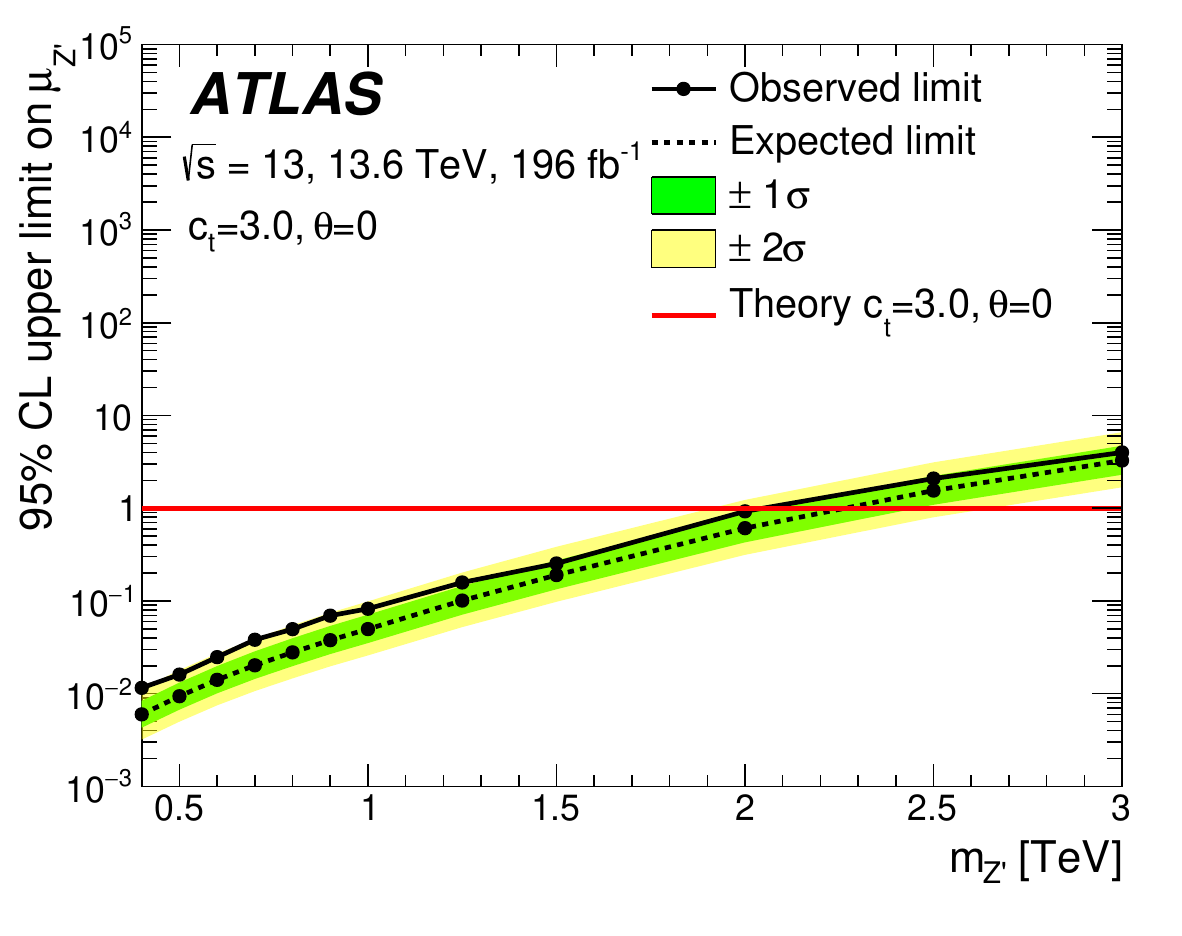}}
\subfloat[]{\includegraphics[width=0.49\linewidth]{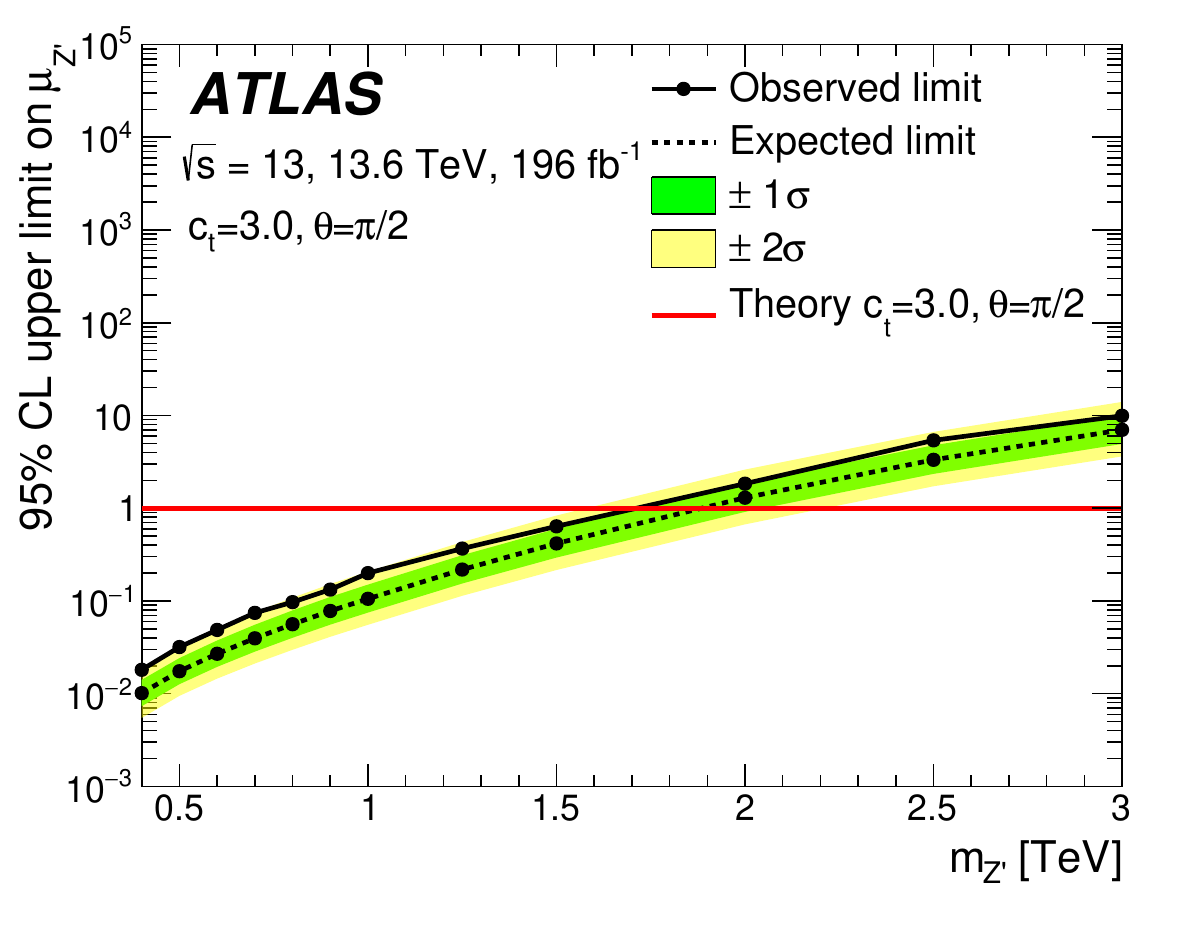}}\\
\caption{\label{fig:limit1dCombined}Observed and expected $95$\% CL upper
limits on the signal strength of the \Zp signal as a function of $m_{\Zp}$ for
(a, b) $c_t = 1$, (c, d) $c_t=3$ and (a,  c) $\theta = 0$
and (b, d) $\theta = \pi/2$. The green (yellow) band shows
the $\pm 1\sigma$ ($\pm 2\sigma$) interval around the expected
limits. The horizontal red lines indicates the theoretical prediction, corresponding to a signal strength of one.}
\end{center}
\end{figure}

\begin{table}[htp]
\caption{ List of uncertainty contributions to the fitted signal strength $\mu$, grouped by category, for the $m_{\Zp}=1000$~\GeV, $c_t=1$, $\theta=0$ signal hypothesis.
The signal strength is defined relative to the predicted \Zp production cross-section.
The contributions from different uncertainty sources are calculated from the covariance matrix of the fitted variables under the Gaussian approximation.}

\begin{center}
\begin{tabular}{l  c  c  }
\hline
Uncertainty source & \multicolumn{2}{c}{$\Delta\mu$}  \\
\hline
\textbf{Signal modelling} & \multicolumn{2}{c}{}  \\
$\tZp, \ttZp (\rightarrow t\bar t)$ &  $+0.001$ & $-0.001$ \\
\hline
\textbf{Background modelling} & \multicolumn{2}{c}{}  \\

$t\bar t t\bar t$  & $+0.069$ & $-0.077$  \\
$t\bar tW$  & $+0.023$ & $-0.025$  \\
$t\bar tt$  & $+0.008$ & $-0.009$ \\
Reducible background  & $+0.008$ & $-0.007$ \\
$t\bar tZ$    & $+0.009$ & $-0.009$ \\
$t\bar tH$ &  $+0.010$ & $-0.011$ \\
Other background  &  $+0.016$ & $-0.016$ \\
\hline
\textbf{Experimental uncertainty} & \multicolumn{2}{c}{}  \\

Jet uncertainties &  $+0.034$ & $-0.035$ \\
Jet flavour tagging ($b$-jets) &   $+0.020$ & $-0.020$ \\
Jet flavour tagging (light-flavour jets) &   $+0.010$ & $-0.009$ \\
Jet flavour tagging ($c$-jets) &   $+0.003$ & $-0.004$ \\
Luminosity &  $+0.006$ & $-0.007$ \\
Other experimental uncertainties &  $+0.011$ & $-0.012$ \\
\hline
\textbf{MC statistical uncertainty} & \multicolumn{2}{c}{}  \\
Simulation sample size &  $+0.014$ & $-0.015$ \\
\hline
Total systematic uncertainty & $+0.089$ & $-0.097$ \\
\hline
\textbf{Statistical} & & \\

HF, Mat. Conv., and Low $m_{\gamma*}$ normalisation & $+0.004$ & $-0.003$ \\
$t\bar tW$ QCD normalisation & $+0.0001$ & $-0.0001$ \\
\hline
Total statistical uncertainty & $+0.20$ & $-0.18$ \\
\hline
\hline

Total uncertainty  &  $+0.22$ & $-0.21$ \\
\hline

\end{tabular}
\label{tab:groupedSyst}
\end{center}
\end{table}%

The dependence on the chirality angle $\theta$ is also investigated.
This parameter quantifies the relative contributions of the \ttZp and \tZp processes, which have
similar experimental signatures despite \tZp involving one fewer top
quark. The likelihood fit therefore yields an anti-correlation between
the two processes, which is reflected in the dependence on the
chirality angle. Figure~\ref{fig:limit2d} shows the two-dimensional scan in the
($c_t^2$, $\theta$) plane, derived from a likelihood fit in which the \ttZp and \tZp signal strengths are
treated as free parameters, at mass points of $m_{\Zp} = 1000$~\GeV and $2000$~\GeV.
The \ttZp production cross-section scales with $c_t^2$ and is independent of $\theta$,
whereas the \tZp contribution depends on both $c_t$ and $\theta$.
As $\theta$ decreases from $\pi/2$ to 0, the signal composition
transitions from purely right-handed four-top production to a mixture
of three-top and four-top processes (left-handed and
right-handed). The addition of three-top events results in more
stringent limits on \Zp~boson production for smaller chirality angles
seen also in Figure~\ref{fig:limit1dCombined}.

\begin{figure}[t]
\begin{center}
\subfloat[]{\includegraphics[width=0.54\linewidth]{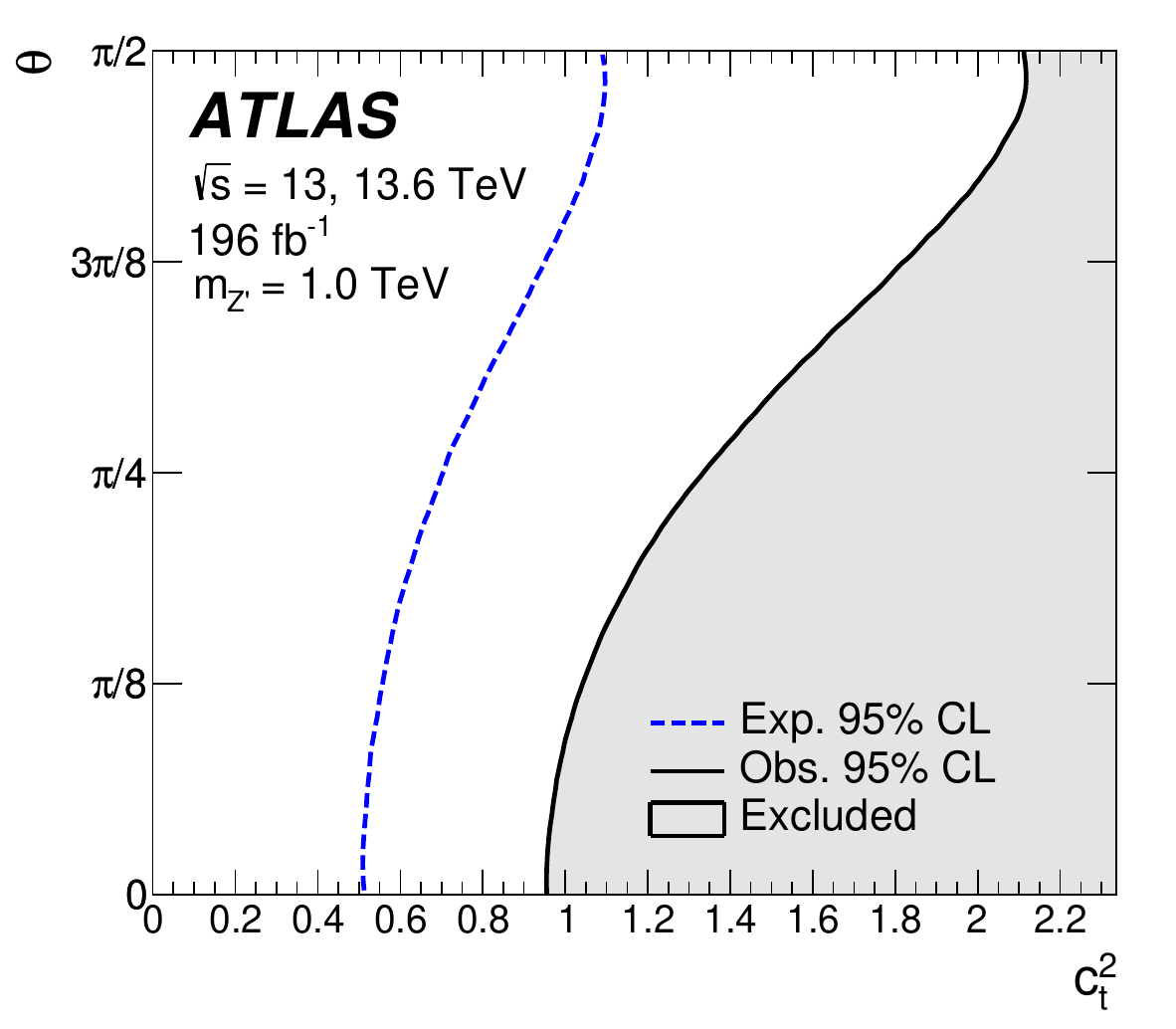}}
\subfloat[]{\includegraphics[width=0.54\linewidth]{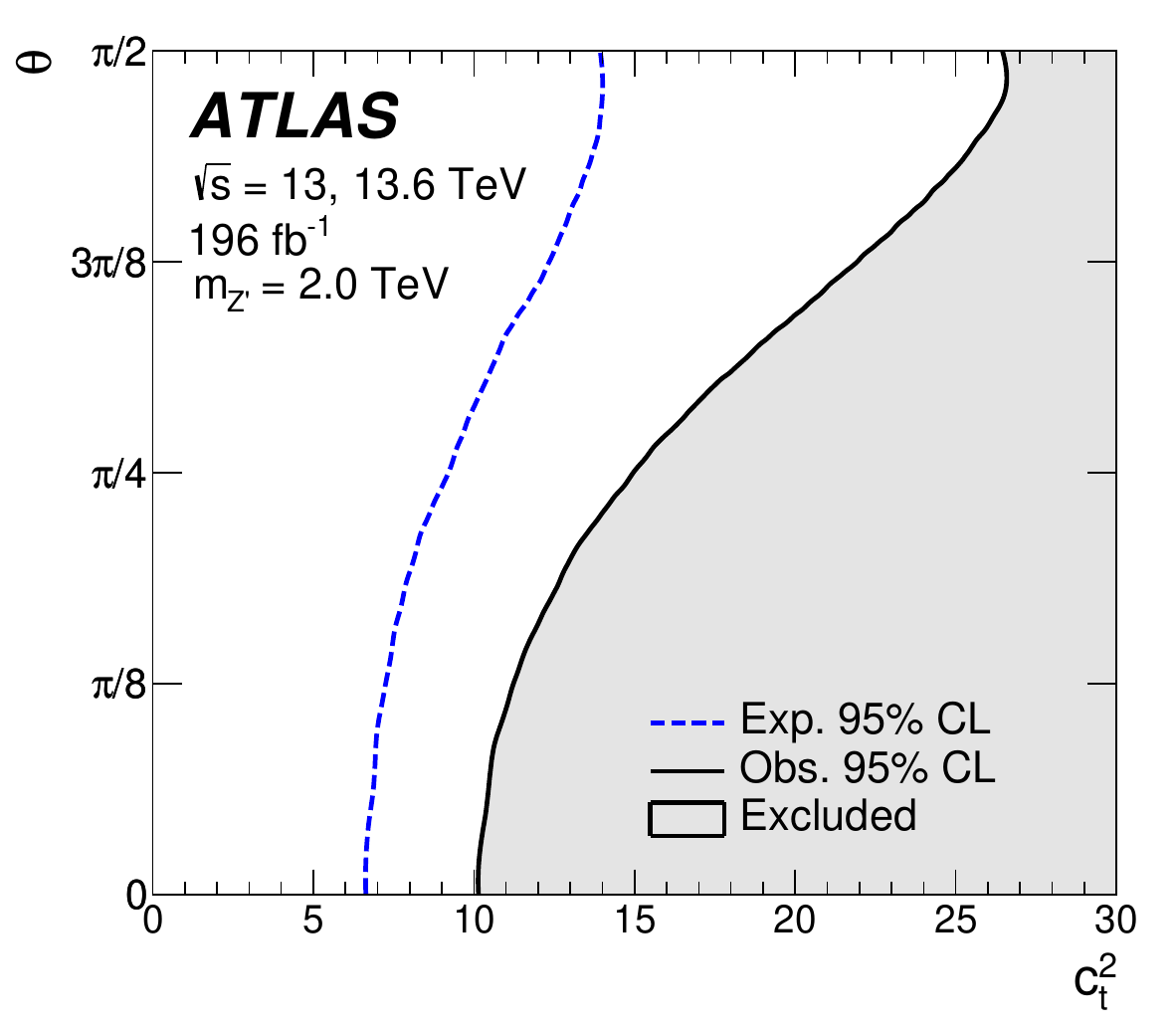}}
\caption{\label{fig:limit2d} Observed and expected 95\% CL contours
in the ($c_t^2$, $\theta$) plane for (a) $m_{\Zp} = 1000$~\GeV and
(b) $m_{\Zp} = 2000$~\GeV, derived from a two-dimensional negative log-likelihood scan.
The grey-shaded region is excluded by the observed 95\% CL limit.
The contour shown in (a) is based on the $c_{t} = 1$ signal sample with a resonance width of 4\%, while the contour in (b) is based on the $c_{t} = 3$ signal sample with a resonance width of 35\%.
}
\end{center}
\end{figure}

% End of text imported from the .//sections/7-results.tex input file

%
\FloatBarrier
\section{Conclusion}
\label{sec:conclusion}

% The next lines are included from the .//sections/8-conclusion.tex input file
A search for a new top-philic colour-singlet heavy resonance in the
process $\Zp \rightarrow t\bar{t}$ in association with either a single top quark or a top-quark pair is presented.
The analysis uses 140~fb$^{-1}$ of proton--proton collision data collected by the
ATLAS detector at the LHC at a centre-of-mass energy of 13~\TeV, and
56~\ifb collected at 13.6~\TeV.
Events with either two same-sign leptons or at least three leptons are
selected. The signal is separated from the backgrounds with a
transformer-based neural network.
Control regions are defined to constrain the dominant $\ttW$ and non-prompt lepton backgrounds.
All regions for both data sets are finally combined in the profile likelihood fit.
No significant excess of data over the Standard Model prediction is observed.
For a \Zp~boson with coupling strength $c_t = 1$ and chirality
angle $\theta = 0$, \Zp~boson masses below 1150~\GeV are excluded at
95\% confidence level.
Limits are also obtained in the $(c_t^2, \theta)$ parameter space, probing
the interplay between the coupling strength and the chirality
structure of the model.
This is a significant improvement in the sensitivity to top-philic \Zp
production compared to ATLAS and CMS results with Run~2
data~\cite{EXOT-2022-14, CMS-B2G-24-009} thanks to the addition of
Run~3 data, improved selection and object performance and the use of
the tranformer-based neural network.

% End of text imported from the .//sections/8-conclusion.tex input file

%
%
%
%
%
\section*{Acknowledgements}
%

%

% The next lines are included from the .//acknowledgements/Acknowledgements.tex input file
%
%

%
%

We thank CERN for the very successful operation of the LHC and its injectors, as well as the support staff at
CERN and at our institutions worldwide without whom ATLAS could not be operated efficiently.

The crucial computing support from all WLCG partners is acknowledged gratefully, in particular from CERN, the ATLAS Tier-1 facilities at TRIUMF/SFU (Canada), NDGF (Denmark, Norway, Sweden), CC-IN2P3 (France), KIT/GridKA (Germany), INFN-CNAF (Italy), NL-T1 (Netherlands), PIC (Spain), RAL (UK) and BNL (USA), the Tier-2 facilities worldwide and large non-WLCG resource providers. Major contributors of computing resources are listed in Ref.~\cite{ATL-SOFT-PUB-2026-001}.

We gratefully acknowledge the support of ANPCyT, Argentina; YerPhI, Armenia; ARC, Australia; BMWFW and FWF, Austria; ANAS, Azerbaijan; CNPq and FAPESP, Brazil; NSERC, NRC and CFI, Canada; CERN; ANID, Chile; CAS, MOST, NSFC and FRFCU, China; Minciencias, Colombia; MEYS CR, Czech Republic; DNRF and DNSRC, Denmark; IN2P3-CNRS and CEA-DRF/IRFU, France; SRNSFG, Georgia; BMFTR, HGF and MPG, Germany; GSRI, Greece; RGC and Hong Kong SAR, China; ICHEP and Academy of Sciences and Humanities, Israel; INFN, Italy; MEXT and JSPS, Japan; CNRST, Morocco; NWO, Netherlands; RCN, Norway; MNiSW, Poland; FCT, Portugal; MNE/IFA, Romania; MSTDI, Serbia; MSSR, Slovakia; ARIS and MVZI, Slovenia; DSI/NRF, South Africa; MICIU/AEI, Spain; SRC and Wallenberg Foundation, Sweden; SERI, SNSF and Cantons of Bern and Geneva, Switzerland; NSTC, Taipei; TENMAK, T\"urkiye; STFC/UKRI, United Kingdom; DOE and NSF, United States of America.

Individual groups and members have received support from BCKDF, CANARIE, CRC and DRAC, Canada; CERN-CZ and PRIMUS, Czech Republic; COST, ERC, ERDF, Horizon 2020 and Marie Sk{\l}odowska-Curie Actions, European Union; Investissements d'Avenir Labex, Investissements d'Avenir Idex and ANR, France; DFG and AvH Foundation, Germany; Herakleitos, Thales and Aristeia programmes co-financed by EU-ESF and the Greek NSRF, Greece; BSF-NSF and MINERVA, Israel; NCN and NAWA, Poland; La Caixa Banking Foundation, CERCA and AGAUR programs from Generalitat de Catalunya and PROMETEO and GenT Programmes Generalitat Valenciana, Spain; G\"{o}ran Gustafssons Stiftelse, Sweden; The Royal Society and Leverhulme Trust, United Kingdom; Eric and Wendy Schmidt Fund for Strategic Innovation, United States of America.

In addition, individual members wish to acknowledge support from Chile: Agencia Nacional de Investigaci\'on y Desarrollo (ANID FONDECYT reg. 1230987, FONDECYT 1230812, FONDECYT 1240864, Fondecyt 3240661, Fondecyt Regular 1240721); China: Fundamental Research Funds for the Central Universities (010-63263105), Chinese Ministry of Science and Technology (MOST-2023YFA1605700, MOST-2023YFA1609300), National Natural Science Foundation of China (NSFC 12275265, NSFC-W2543005); Czech Republic: Czech Science Foundation (GACR - 24-11373S), Ministry of Education Youth and Sports (ERC-CZ-LL2327, FORTE CZ.02.01.01/00/22\_008/0004632); EU: H2020 European Research Council (ERC - 101002463); European Union: European Research Council (BARD No. 101116429, ERC - 101219398, ERC 101089007), European Regional Development Fund (HE COFUND GA No.101081355, ERDF), Marie Sklodowska-Curie Actions (GAP-101168829); France: Agence Nationale de la Recherche (ANR-22-EDIR-0002, ANR-24-CE31-0504-01); Germany: Deutsche Forschungsgemeinschaft (DFG - 469666862); China: Research Grants Council (GRF); Italy: Istituto Nazionale di Fisica Nucleare (LHC-MIUR - 28003/2025), Ministero dell'Università e della Ricerca (NextGenEU I53D23000820006 M4C2.1.1, SOE2024\_0000023); Japan: Japan Society for the Promotion of Science (JSPS KAKENHI  JP25H0063, JSPS KAKENHI JP22H04944, JSPS KAKENHI JP24K23939, JSPS KAKENHI JP24KK0251, JSPS KAKENHI JP25H00650, JSPS KAKENHI JP25H01291, JSPS KAKENHI JP25K01011, JSPS KAKENHI JP25K01023, JSPS KAKENHI JP25KK0047); Poland: Polish National Science Centre (NCN 2021/42/E/ST2/00350, NCN OPUS 2023/51/B/ST2/02507, NCN OPUS nr 2022/47/B/ST2/03059, NCN UMO-2019/34/E/ST2/00393, UMO-2022/47/O/ST2/00148, UMO-2023/49/B/ST2/04085, UMO-2023/51/B/ST2/00920, UMO-2024/53/N/ST2/00869); Spain: Agència de Gestió d'Ajuts Universitaris i de Recerca. (AGAUR - 2023 BP 00141), Ministry of Science and Innovation (RYC2019-028510-I, RYC2020-030254-I, RYC2021-031273-I, RYC2022-038164-I), Ministerio de Ciencia, Innovación y Universidades/Agencia Estatal de Investigaci\'on (EU NextGenerationEU (PRTR-C17.I1), PID2022-142604OB-C22); Sweden: Carl Trygger Foundation (Carl Trygger Foundation CTS 22:2312), Swedish Research Council (Swedish Research Council 2023-04654, VR 2021-03651, VR 2022-03845, VR 2022-04683, VR 2023-03403, VR 2024-05451, VR 2025-05940), Knut and Alice Wallenberg Foundation (KAW 2023.0366); United Kingdom: The Binks Trust, Royal Society (NIF-R1-231091); United States of America: U.S. Department of Energy (ECA DE-AC02-76SF00515), John Templeton Foundation (John Templeton Foundation 63206), Neubauer Family Foundation.

%
%

% End of text imported from the .//acknowledgements/Acknowledgements.tex input file

%
%
%
%
%
%

%
%

%
%
%
\printbibliography

@article{CDF:1995wbb,
    author   = "{CDF Collaboration}",
    title = "{Observation of top quark production in $\bar{p}p$ collisions with the Collider Detector at Fermilab}",
    eprint = "hep-ex/9503002",
    archivePrefix = "arXiv",
    reportNumber = "FERMILAB-PUB-95-022-E, CDF-PUB-TOP-PUBLIC-3040, ANL-HEP-PR-95-44",
    doi = "10.1103/PhysRevLett.74.2626",
    journal = "Phys. Rev. Lett.",
    volume = "74",
    pages = "2626--2631",
    year = "1995"
}

@article{D0:1995jca,
    author = "{D0 Collaboration}",
    title = "{Observation of the top quark}",
    eprint = "hep-ex/9503003",
    archivePrefix = "arXiv",
    reportNumber = "FERMILAB-PUB-95-028-E, D0-2698, D0-PUB-95-7",
    doi = "10.1103/PhysRevLett.74.2632",
    journal = "Phys. Rev. Lett.",
    volume = "74",
    pages = "2632--2637",
    year = "1995"
}

@article{Susskind:1978ms,
    author = "Susskind, Leonard",
    title = "{Dynamics of Spontaneous Symmetry Breaking in the Weinberg-Salam Theory}",
    reportNumber = "SLAC-PUB-2142",
    doi = "10.1103/PhysRevD.20.2619",
    journal = "Phys. Rev. D",
    volume = "20",
    pages = "2619--2625",
    year = "1979"
}

@article{Ferretti:2013kya,
    author = "Ferretti, Gabriele and Karateev, Denis",
    title = "{Fermionic UV completions of Composite Higgs models}",
    eprint = "1312.5330",
    archivePrefix = "arXiv",
    primaryClass = "hep-ph",
    doi = "10.1007/JHEP03(2014)077",
    journal = "JHEP",
    volume = "03",
    pages = "077",
    year = "2014"
}

@article{Vecchi:2015fma,
    author = "Vecchi, Luca",
    title = "{A dangerous irrelevant UV-completion of the composite Higgs}",
    eprint = "1506.00623",
    archivePrefix = "arXiv",
    primaryClass = "hep-ph",
    doi = "10.1007/JHEP02(2017)094",
    journal = "JHEP",
    volume = "02",
    pages = "094",
    year = "2017"
}

@article{Agashe:2003zs,
    author = "Agashe, Kaustubh and Delgado, Antonio and May, Michael J. and Sundrum, Raman",
    title = "{RS1, custodial isospin and precision tests}",
    eprint = "hep-ph/0308036",
    archivePrefix = "arXiv",
    doi = "10.1088/1126-6708/2003/08/050",
    journal = "JHEP",
    volume = "08",
    pages = "050",
    year = "2003"
}

@article{Agashe:2004rs,
    author = "Agashe, Kaustubh and Contino, Roberto and Pomarol, Alex",
    title = "{The Minimal composite Higgs model}",
    eprint = "hep-ph/0412089",
    archivePrefix = "arXiv",
    reportNumber = "UAB-FT-567",
    doi = "10.1016/j.nuclphysb.2005.04.035",
    journal = "Nucl. Phys. B",
    volume = "719",
    pages = "165--187",
    year = "2005"
}

@article{CMS-B2G-24-009,
      title       =  "{Search for a top-philic Z' boson decaying into a $\mathrm{t\bar{t}}$ pair in a final state with jets and an electron or muon in proton-proton collisions at $\sqrt{s}$ = 13.6 TeV}",
      author      = "{CMS Collaboration}", 
      year        =  "2026",
      eprint      =  "2608.01251",
      archivePrefix = "arXiv",
      primaryClass  = "hep-ex",
      journal = "submitted to JHEP"
}

@article{salt2025,
  author = {Jackson Barr and Diptaparna Biswas and Maxence Draguet and Philipp Gadow and Emil Haines and Osama Karkout and Dmitrii Kobylianskii and Wei Sheng Lai and Matthew Leigh and Nicholas Luongo and Ivan Oleksiyuk and Nikita Pond and Sébastien Rettie and Andrius Vaitkus and Samuel Van Stroud and Johannes Wagner},
  title = {Salt: Multimodal Multitask Machine Learning for High Energy Physics},
  journal = {Journal of Open Source Software},
  volume = {10},
  year = {2025},
  doi = {10.21105/joss.07217},
}

@misc{vaswani2023at,
      title={Attention Is All You Need}, 
      author={Ashish Vaswani and Noam Shazeer and Niki Parmar and Jakob Uszkoreit and Llion Jones and Aidan N. Gomez and Lukasz Kaiser and Illia Polosukhin},
      year={2023},
      eprint={1706.03762},
      archivePrefix={arXiv},
      primaryClass={cs.CL},
}

@article{Pinto:2023yob,
    author = "Pinto, Andr{\'e}s and Wu, Zhibo and Balli, Fabrice and Berger, Nicolas and Boonekamp, Maarten and Chapon, {\'E}milien and Kawamoto, Tatsuo and Malaescu, Bogdan",
    title = "{Uncertainty components in profile likelihood fits}",
    eprint = "2307.04007",
    archivePrefix = "arXiv",
    primaryClass = "physics.data-an",
    doi = "10.1140/epjc/s10052-024-12877-5",
    journal = "Eur. Phys. J. C",
    volume = "84",
    number = "6",
    pages = "593",
    year = "2024"
}

@Article{Kim_2016,
  title = {Probing TeV scale top-philic resonances with boosted top-tagging at the high luminosity LHC},
  author = {Kim, Jeong Han and Kong, Kyoungchul and Lee, Seung J. and Mohlabeng, Gopolang},
  journal = {Phys. Rev. D},
  volume = {94},
  pages = {035023},
  numpages = {20},
  year = {2016},
  month = {8},
  doi = {10.1103/PhysRevD.94.035023},
  archivePrefix = "arXiv",
  eprint = "1604.07421"
}

@Article{Melissa:4top,
   author        = {Melissa van Beekveld and Anna Kulesza and Michele Lupattelli and Tommaso Saracco},
   title         = {Invariant-mass threshold resummation for the production of four top quarks at the LHC},
   journal       = {JHEP},
   volume        = {2025},
   number        = {10},
   pages         = {209},
   year          = {2025},
   doi           = {10.1007/JHEP10(2025)209},
   archivePrefix = "arXiv",
   eprint        = {2505.10381},
   primaryClass  = {hep-ph},
}

@article{Gutschow:2018tuk,
    author = {G\"utschow, Christian and Lindert, Jonas M. and Sch\"onherr, Marek},
    title = "{Multi-jet merged top-pair production including electroweak corrections}",
    eprint = "1803.00950",
    archivePrefix = "arXiv",
    primaryClass = "hep-ph",
    reportNumber = "CERN-TH-18-039, CERN-TH-2018-039, IPPP/18/13, MCnet-18-05, IPPP-18-13, MCNET-18-05",
    doi = "10.1140/epjc/s10052-018-5804-2",
    journal = "Eur. Phys. J. C",
    volume = "78",
    number = "4",
    pages = "317",
    year = "2018"
}

@article{Cacciari:2008,
    author = "Cacciari, Matteo and Salam, Gavin P. and Soyez, Gregory",
    title = "{The catchment area of jets}",
    doi = "10.1088/1126-6708/2008/04/005",
    journal = "JHEP",
    volume = "04",
    pages = "005",
    year = "2008",
    eprint  = "0802.1188",
    archivePrefix = {arXiv},
    primaryClass  = {hep-ph},
}

@article{Mattelaer:2016gcx,
    author = "Mattelaer, Olivier",
    title = "{On the maximal use of Monte Carlo samples: re-weighting events at NLO accuracy}",
    eprint = "1607.00763",
    archivePrefix = "arXiv",
    primaryClass = "hep-ph",
    reportNumber = "IPPP-16-55, MCNET-16-25",
    doi = "10.1140/epjc/s10052-016-4533-7",
    journal = "Eur. Phys. J. C",
    volume = "76",
    number = "12",
    pages = "674",
    year = "2016"
}

@Article{TOPQ-2021-08-witherratum,
    author         = "{ATLAS Collaboration}",
    title          = "{Observation of four-top-quark production in the multilepton final state with the ATLAS detector}",
    journal        = "Eur. Phys. J. C",
    volume         = "83",
    year           = "2023",
    pages          = "496",
    doi            = "10.1140/epjc/s10052-023-11573-0",
    reportNumber   = "CERN-EP-2023-055",
    eprint         = "2303.15061",
    archivePrefix  = "arXiv",
    primaryClass   = "hep-ex",
    addendum       = "[Erratum: \href{https://doi.org/10.1140/epjc/s10052-024-12458-6}{Eur. Phys. J. C \textbf{84} (2024) 156}]",
}

@Article{EXOT-2019-26-witherratum,
    author         = "{ATLAS Collaboration}",
    title          = "{Search for \(t\bar{t}H/A \rightarrow t\bar{t} t\bar{t}\) production in the multilepton final state in proton--proton collisions at \(\sqrt{s} = 13\,\text{TeV}\) with the ATLAS detector}",
    journal        = "JHEP",
    volume         = "07",
    year           = "2023",
    pages          = "203",
    doi            = "10.1007/JHEP07(2023)203",
    reportNumber   = "CERN-EP-2022-170",
    eprint         = "2211.01136",
    archivePrefix  = "arXiv",
    primaryClass   = "hep-ex",
    addendum       = "[Erratum: \href{https://doi.org/10.1007/JHEP05(2025)159}{JHEP \textbf{05} (2025) 159}]",
}

@Article{Cacciari:2008gp,
     author    = "Cacciari, Matteo and Salam, Gavin P. and Soyez, Gregory",
     title     = "{The anti-\(k_{t}\) jet clustering algorithm}",
     journal   = "JHEP",
     volume    = "04",
     year      = "2008",
     pages     = "063",
     eprint    = "0802.1189",
     archivePrefix = "arXiv",
     primaryClass  =  "hep-ph",
     doi       = "10.1088/1126-6708/2008/04/063",
     SLACcitation  = "%%CITATION = 0802.1189;%%"
}

@Article{Fastjet,
      author         = "Cacciari, Matteo and Salam, Gavin P. and Soyez, Gregory",
      title          = "{FastJet user manual}",
      journal        = "Eur. Phys. J. C",
      volume         = "72",
      year           = "2012",
      pages          = "1896",
      doi            = "10.1140/epjc/s10052-012-1896-2",
      eprint         = "1111.6097",
      archivePrefix  = "arXiv",
      primaryClass   = "hep-ph",
      reportNumber   = "CERN-PH-TH-2011-297",
      SLACcitation   = "%%CITATION = ARXIV:1111.6097;%%"
}

@Article{Butterworth:2015oua,
      author         = "Butterworth, Jon and others",
      title          = "{PDF4LHC recommendations for LHC Run II}",
      journal        = "J. Phys. G",
      volume         = "43",
      year           = "2016",
      pages          = "023001",
      doi            = "10.1088/0954-3899/43/2/023001",
      eprint         = "1510.03865",
      archivePrefix  = "arXiv",
      primaryClass   = "hep-ph",
      reportNumber   = "OUTP-15-17P, SMU-HEP-15-12, TIF-UNIMI-2015-14,
                        LCTS-2015-27, CERN-PH-TH-2015-249",
      SLACcitation   = "%%CITATION = ARXIV:1510.03865;%%"
}

@article{Ball:2014uwa,
      author         = "{NNPDF Collaboration} and Ball, Richard D. and others",
      title          = "{Parton distributions for the LHC run II}",
      journal        = "JHEP",
      volume         = "04",
      year           = "2015",
      pages          = "040",
      doi            = "10.1007/JHEP04(2015)040",
      eprint         = "1410.8849",
      archivePrefix  = "arXiv",
      primaryClass   = "hep-ph",
      reportNumber   = "EDINBURGH-2014-15, IFUM-1034-FT, CERN-PH-TH-2013-253,
                        OUTP-14-11P, CAVENDISH-HEP-14-11",
      SLACcitation   = "%%CITATION = ARXIV:1410.8849;%%"
}

@article{Sjostrand:2014zea,
      author         = "Sj{\"o}strand, Torbj{\"o}rn and Ask, Stefan and Christiansen,
                        Jesper R. and Corke, Richard and Desai, Nishita and Ilten,
                        Philip and Mrenna, Stephen and Prestel, Stefan and
                        Rasmussen, Christine O. and Skands, Peter Z.",
      title          = "{An introduction to PYTHIA 8.2}",
      journal        = "Comput. Phys. Commun.",
      volume         = "191",
      year           = "2015",
      pages          = "159",
      doi            = "10.1016/j.cpc.2015.01.024",
      eprint         = "1410.3012",
      archivePrefix  = "arXiv",
      primaryClass   = "hep-ph",
      reportNumber   = "LU-TP-14-36, MCNET-14-22, CERN-PH-TH-2014-190,
                        FERMILAB-PUB-14-316-CD, DESY-14-178, SLAC-PUB-16122,
                        --FERMILAB-PUB-14-316-CD",
      SLACcitation   = "%%CITATION = ARXIV:1410.3012;%%"
}

@Article{Bierlich:2022pfr,
    author        = {Bierlich, Christian and Chakraborty, Smita and Desai, Nishita and Gellersen, Leif and Helenius, Ilkka and Ilten, Philip and Lönnblad, Leif and Mrenna, Stephen and Prestel, Stefan and Preuss, Christian T. and Sjöstrand, Torbjörn and Skands, Peter and Utheim, Marius and Verheyen, Rob},
    title         = "{A comprehensive guide to the physics and usage of PYTHIA 8.3}",
    journal       = {SciPost Phys. Codebases},
    pages         = {8},
    year          = {2022},
    publisher     = {SciPost},
    doi           = {10.21468/SciPostPhysCodeb.8},
    eprint        = "2203.11601",
    archivePrefix = "arXiv",
    primaryClass  = "hep-ph",
    reportNumber  = "LU-TP 22-16, MCNET-22-04, FERMILAB-PUB-22-227-SCD",
}

@Article{Frixione:2008yi,
     author    = "Frixione, Stefano and Laenen, Eric and Motylinski, Patrick and White, Chris and Webber, Bryan R.",
     title     = "{Single-top hadroproduction in association with a \(W\) boson}",
     journal   = "JHEP",
     volume    = "07",
     year      = "2008",
     pages     = "029",
     eprint    = "0805.3067",
     archivePrefix = "arXiv",
     primaryClass  =  "hep-ph",
     doi       = "10.1088/1126-6708/2008/07/029"
}

@Article{Alwall:2011uj,
      author         = "Alwall, Johan and Herquet, Michel and Maltoni, Fabio and
                        Mattelaer, Olivier and Stelzer, Tim",
      title          = "{MadGraph 5: going beyond}",
      journal        = "JHEP",
      volume         = "06",
      pages          = "128",
      doi            = "10.1007/JHEP06(2011)128",
      year           = "2011",
      eprint         = "1106.0522",
      archivePrefix  = "arXiv",
      primaryClass   = "hep-ph"
}

@Article{Alwall:2014hca,
      author         = "Alwall, J. and Frederix, R. and Frixione, S. and Hirschi,
                        V. and Maltoni, F. and Mattelaer, O. and Shao, H. -S. and
                        Stelzer, T. and Torrielli, P. and Zaro, M.",
      title          = "{The automated computation of tree-level and
                        next-to-leading order differential cross sections, and
                        their matching to parton shower simulations}",
      journal        = "JHEP",
      volume         = "07",
      year           = "2014",
      pages          = "079",
      doi            = "10.1007/JHEP07(2014)079",
      eprint         = "1405.0301",
      archivePrefix  = "arXiv",
      primaryClass   = "hep-ph",
      reportNumber   = "CERN-PH-TH-2014-064, CP3-14-18, LPN14-066, MCNET-14-09,
                        ZU-TH-14-14",
      SLACcitation   = "%%CITATION = ARXIV:1405.0301;%%"
}

@article{Frederix:2012ps,
      author         = "Frederix, Rikkert and Frixione, Stefano",
      title          = "{Merging meets matching in MC@NLO}",
      journal        = "JHEP",
      volume         = "12",
      year           = "2012",
      pages          = "061",
      doi            = "10.1007/JHEP12(2012)061",
      eprint         = "1209.6215",
      archivePrefix  = "arXiv",
      primaryClass   = "hep-ph",
      reportNumber   = "CERN-PH-TH-2012-247, ZU-TH-21-12",
      SLACcitation   = "%%CITATION = ARXIV:1209.6215;%%"
}

@article{Artoisenet:2012st,
      author         = "Artoisenet, Pierre and Frederix, Rikkert and Mattelaer,
                        Olivier and Rietkerk, Robbert",
      title          = "{Automatic spin-entangled decays of heavy resonances in
                        Monte Carlo simulations}",
      journal        = "JHEP",
      volume         = "03",
      pages          = "015",
      doi            = "10.1007/JHEP03(2013)015",
      year           = "2013",
      eprint         = "1212.3460",
      archivePrefix  = "arXiv",
      primaryClass   = "hep-ph",
      reportNumber   = "NIKHEF-2012-021, CERN-PH-TH-2012-329",
      SLACcitation   = "%%CITATION = ARXIV:1212.3460;%%",
}

@article{Frixione:2007zp,
      author         = "Frixione, Stefano and Laenen, Eric and Motylinski,
                        Patrick and Webber, Bryan R.",
      title          = "{Angular correlations of lepton pairs from vector boson
                        and top quark decays in Monte Carlo simulations}",
      journal        = "JHEP",
      volume         = "04",
      year           = "2007",
      pages          = "081",
      doi            = "10.1088/1126-6708/2007/04/081",
      eprint         = "hep-ph/0702198",
      archivePrefix  = "arXiv",
      reportNumber   = "CAVENDISH-HEP-07-01, GEF-TH-09-2007, ITP-UU-07-10,
                        NIKHEF-2007-004",
      SLACcitation   = "%%CITATION = HEP-PH/0702198;%%"
}

@Article{Bahr:2008pv,
      author         = "B{\"a}hr, M. and others",
      title          = "{Herwig++ physics and manual}",
      journal        = "Eur. Phys. J. C",
      volume         = "58",
      year           = "2008",
      pages          = "639",
      doi            = "10.1140/epjc/s10052-008-0798-9",
      eprint         = "0803.0883",
      archivePrefix  = "arXiv",
      primaryClass   = "hep-ph",
      reportNumber   = "CERN-PH-TH-2008-038, CAVENDISH-HEP-08-03, KA-TP-05-2008,
                        DCPT-08-22, IPPP-08-11, CP3-08-05",
      SLACcitation   = "%%CITATION = ARXIV:0803.0883;%%"
}

@Article{Bellm:2015jjp,
      author         = "Bellm, Johannes and others",
      title          = "{Herwig 7.0/Herwig++ 3.0 release note}",
      journal        = "Eur. Phys. J. C",
      volume         = "76",
      year           = "2016",
      number         = "4",
      pages          = "196",
      doi            = "10.1140/epjc/s10052-016-4018-8",
      eprint         = "1512.01178",
      archivePrefix  = "arXiv",
      primaryClass   = "hep-ph",
      reportNumber   = "CERN-PH-TH-2015-289, MAN-HEP-2015-15, IFJPAN-IV-2015-13,
                        HERWIG-2015-01, KA-TP-18-2015, DCPT-15-142, MCNET-15-28,
                        IPPP-15-71, --HERWIG-2015-01",
      SLACcitation   = "%%CITATION = ARXIV:1512.01178;%%"
}

@Article{Bellm:2019zci,
    author = "Bellm, Johannes and others",
    title = "{Herwig 7.2 release note}",
    eprint = "1912.06509",
    archivePrefix = "arXiv",
    primaryClass = "hep-ph",
    reportNumber = "MAN/HEP/2019/011, CERN-TH-2019-213, IFJPAN-IV-2019-18, HERWIG-2019-02, UWTHPH-19-36, KA-TP-24-2019, LU-TP 19-57, MCnet-19-28, IPPP/19/91",
    doi = "10.1140/epjc/s10052-020-8011-x",
    journal = "Eur. Phys. J. C",
    volume = "80",
    pages = "452",
    year = "2020"
}

@Article{Nason:2004rx,
      author         = "Nason, Paolo",
      title          = "{A new method for combining NLO QCD with shower Monte Carlo algorithms}",
      journal        = "JHEP",
      volume         = "11",
      pages          = "040",
      doi            = "10.1088/1126-6708/2004/11/040",
      year           = "2004",
      eprint         = "hep-ph/0409146",
      archivePrefix  = "arXiv",
}

@Article{Frixione:2007vw,
      author         = "Frixione, Stefano and Nason, Paolo and Oleari, Carlo",
      title          = "{Matching NLO QCD computations with parton shower
                        simulations: the POWHEG method}",
      journal        = "JHEP",
      volume         = "11",
      pages          = "070",
      doi            = "10.1088/1126-6708/2007/11/070",
      year           = "2007",
      eprint         = "0709.2092",
      archivePrefix  = "arXiv",
      primaryClass   = "hep-ph",
}

@Article{Alioli:2010xd,
      author         = "Alioli, Simone and Nason, Paolo and Oleari, Carlo and Re,
                        Emanuele",
      title          = "{A general framework for implementing NLO calculations in
                        shower Monte Carlo programs: the POWHEG BOX}",
      journal        = "JHEP",
      volume         = "06",
      pages          = "043",
      doi            = "10.1007/JHEP06(2010)043",
      year           = "2010",
      eprint         = "1002.2581",
      archivePrefix  = "arXiv",
      primaryClass   = "hep-ph",
}

@article{Hartanto:2015uka,
      author         = "Hartanto, Heribertus B. and J{\"a}ger, Barbara and Reina,
                        Laura and Wackeroth, Doreen",
      title          = "{Higgs boson production in association with top quarks in
                        the POWHEG BOX}",
      journal        = "Phys. Rev. D",
      volume         = "91",
      year           = "2015",
      number         = "9",
      pages          = "094003",
      doi            = "10.1103/PhysRevD.91.094003",
      eprint         = "1501.04498",
      archivePrefix  = "arXiv",
      primaryClass   = "hep-ph",
      SLACcitation   = "%%CITATION = ARXIV:1501.04498;%%"
}

@article{Frixione:2007nw,
      author         = "Frixione, Stefano and Ridolfi, Giovanni and Nason, Paolo",
      title          = "{A positive-weight next-to-leading-order Monte Carlo for
                        heavy flavour hadroproduction}",
      journal        = "JHEP",
      volume         = "09",
      pages          = "126",
      doi            = "10.1088/1126-6708/2007/09/126",
      year           = "2007",
      eprint         = "0707.3088",
      archivePrefix  = "arXiv",
      primaryClass   = "hep-ph",
      SLACcitation   = "%%CITATION = ARXIV:0707.3088;%%",
}

@article{Frederix:2017wme,
      author         = "Frederix, Rikkert and Pagani, Davide and Zaro, Marco",
      title          = "{Large NLO corrections in \(t\bar{t}W^{\pm}\) and
                        \(t\bar{t}t\bar{t}\) hadroproduction from supposedly
                        subleading EW contributions}",
      journal        = "JHEP",
      volume         = "02",
      year           = "2018",
      pages          = "031",
      doi            = "10.1007/JHEP02(2018)031",
      eprint         = "1711.02116",
      archivePrefix  = "arXiv",
      primaryClass   = "hep-ph",
      SLACcitation   = "%%CITATION = ARXIV:1711.02116;%%"
}

@article{deFlorian:2016spz,
      author         = "de Florian, D. and others",
      title          = "{Handbook of LHC Higgs Cross Sections: 4. Deciphering the Nature of the Higgs Sector}",
      collaboration  = "LHC Higgs Cross Section Working Group",
      doi            = "10.23731/CYRM-2017-002",
      year           = "2017",
      eprint         = "1610.07922",
      archivePrefix  = "arXiv",
      primaryClass   = "hep-ph",
      reportNumber   = "FERMILAB-FN-1025-T, CERN-2017-002-M",
      SLACcitation   = "%%CITATION = ARXIV:1610.07922;%%"
}

@Article{Hoeche:2009rj,
      author         = "H{\"o}che, Stefan and Krauss, Frank and Schumann, Steffen and Siegert, Frank",
      title          = "{QCD matrix elements and truncated showers}",
      journal        = "JHEP",
      volume         = "05",
      pages          = "053",
      doi            = "10.1088/1126-6708/2009/05/053",
      year           = "2009",
      eprint         = "0903.1219",
      archivePrefix  = "arXiv",
      primaryClass   = "hep-ph",
}

@Article{Schumann:2007mg,
      author         = "Schumann, Steffen and Krauss, Frank",
      title          = "{A parton shower algorithm based on Catani--Seymour dipole factorisation}",
      journal        = "JHEP",
      volume         = "03",
      pages          = "038",
      doi            = "10.1088/1126-6708/2008/03/038",
      year           = "2008",
      eprint         = "0709.1027",
      archivePrefix  = "arXiv",
      primaryClass   = "hep-ph",
}

@article{Hoeche:2012yf,
      author         = "H{\"o}che, Stefan and Krauss, Frank and Sch{\"o}nherr, Marek and
                        Siegert, Frank",
      title          = "{QCD matrix elements + parton showers. The NLO case}",
      journal        = "JHEP",
      volume         = "04",
      year           = "2013",
      pages          = "027",
      doi            = "10.1007/JHEP04(2013)027",
      eprint         = "1207.5030",
      archivePrefix  = "arXiv",
      primaryClass   = "hep-ph",
      reportNumber   = "SLAC-PUB-15191, IPPP-12-52, DCPT-12-104, LPN12-081,
                        FR-PHENO-2012-017, MCNET-12-09, --FR-PHENO-2012-017",
      SLACcitation   = "%%CITATION = ARXIV:1207.5030;%%"
}

@article{Bothmann:2019yzt,
      author = "Bothmann, Enrico and others",
      title = "{Event generation with Sherpa 2.2}",
      journal = "SciPost Phys.",
      volume = "7",
      year = "2019",
      number = "3",
      pages = "034",
      doi = "10.21468/SciPostPhys.7.3.034",
      reportNumber = "FERMILAB-PUB-19-218-T, SLAC-PUB-17433, IPPP/19/42, MCNET-19-11",
      eprint = "1905.09127",
      archivePrefix = "arXiv",
      primaryClass = "hep-ph",
}

@article{Hoeche:2011fd,
      author         = "H{\"o}che, Stefan and Krauss, Frank and Sch{\"o}nherr, Marek and
                        Siegert, Frank",
      title          = "{A critical appraisal of NLO+PS matching methods}",
      journal        = "JHEP",
      volume         = "09",
      year           = "2012",
      pages          = "049",
      doi            = "10.1007/JHEP09(2012)049",
      eprint         = "1111.1220",
      archivePrefix  = "arXiv",
      primaryClass   = "hep-ph",
      reportNumber   = "SLAC-PUB-14661, IPPP-11-67, DCPT-11-134, LPN11-58,
                        FR-PHENO-2011-019, MCNET-11-24",
      SLACcitation   = "%%CITATION = ARXIV:1111.1220;%%"
}

@article{Catani:2001cc,
      author         = "Catani, S. and Krauss, F. and Webber, B. R. and Kuhn, R.",
      title          = "{QCD Matrix Elements + Parton Showers}",
      journal        = "JHEP",
      volume         = "11",
      year           = "2001",
      pages          = "063",
      doi            = "10.1088/1126-6708/2001/11/063",
      eprint         = "hep-ph/0109231",
      archivePrefix  = "arXiv",
      reportNumber   = "CERN-TH-2000-367, CAVENDISH-HEP-00-03",
      SLACcitation   = "%%CITATION = HEP-PH/0109231;%%"
}

@article{Cascioli:2011va,
      author         = "Cascioli, Fabio and Maierh{\"o}fer, Philipp and Pozzorini, Stefano",
      title          = "{Scattering Amplitudes with Open Loops}",
      journal        = "Phys. Rev. Lett.",
      volume         = "108",
      year           = "2012",
      pages          = "111601",
      doi            = "10.1103/PhysRevLett.108.111601",
      eprint         = "1111.5206",
      archivePrefix  = "arXiv",
      primaryClass   = "hep-ph",
      reportNumber   = "ZU-TH-23-11, LPN11-66",
      SLACcitation   = "%%CITATION = ARXIV:1111.5206;%%"
}

@article{Buccioni:2019sur,
    author = "Buccioni, Federico and Lang, Jean-Nicolas and Lindert, Jonas M. and Maierh{\"o}fer, Philipp and Pozzorini, Stefano and Zhang, Hantian and Zoller, Max F.",
    title = "{OpenLoops 2}",
    eprint = "1907.13071",
    archivePrefix = "arXiv",
    primaryClass = "hep-ph",
    reportNumber = "IPPP/19/62, FR-PHENO-2019-12, PSI-PR-19-15, ZU-TH 37/19",
    doi = "10.1140/epjc/s10052-019-7306-2",
    journal = "Eur. Phys. J. C",
    volume = "79",
    number = "10",
    pages = "866",
    year = "2019"
}

@article{Denner:2016kdg,
      author         = "Denner, Ansgar and Dittmaier, Stefan and Hofer, Lars",
      title          = "{\textsc{Collier}: A fortran-based complex one-loop library in
                        extended regularizations}",
      journal        = "Comput. Phys. Commun.",
      volume         = "212",
      year           = "2017",
      pages          = "220-238",
      doi            = "10.1016/j.cpc.2016.10.013",
      eprint         = "1604.06792",
      archivePrefix  = "arXiv",
      primaryClass   = "hep-ph",
      reportNumber   = "FR-PHENO-2016-003, ICCUB-16-016",
      SLACcitation   = "%%CITATION = ARXIV:1604.06792;%%"
}

@article{Kallweit:2015dum,
    author = "Kallweit, Stefan and Lindert, Jonas M. and Maierh{\"o}fer, Philipp and Pozzorini, Stefano and Sch{\"o}nherr, Marek",
    title = "{NLO QCD+EW predictions for V + jets including off-shell vector-boson decays and multijet merging}",
    eprint = "1511.08692",
    archivePrefix = "arXiv",
    primaryClass = "hep-ph",
    reportNumber = "DCPT-15-140, FR-PHENO-2015-014, IPPP-15-70, MCNET-15-23, ZU-TH-41-15, MITP-15-108",
    doi = "10.1007/JHEP04(2016)021",
    journal = "JHEP",
    volume = "04",
    pages = "021",
    year = "2016"
}

@article{Agostinelli:2002hh,
      author         = "Agostinelli, S. and others",
      title          = "{\textsc{Geant4} -- a simulation toolkit}",
      journal        = "Nucl. Instrum. Meth. A",
      volume         = "506",
      year           = "2003",
      pages          = "250",
      doi            = "10.1016/S0168-9002(03)01368-8",
      reportNumber   = "SLAC-PUB-9350, FERMILAB-PUB-03-339",
      SLACcitation   = "%%CITATION = NUIMA,A506,250;%%"
}

@software{FastFrames,
  author       = {{ATLAS Collaboration}},
  title        = {FastFrames},
  publisher    = {Zenodo},
  doi          = {10.5281/zenodo.14773274},
  url          = {https://doi.org/10.5281/zenodo.14773274}
}

@software{TRExFitter,
  author       = {{ATLAS Collaboration}},
  title        = {TRExFitter},
  publisher    = {Zenodo},
  doi          = {10.5281/zenodo.14845712},
  url          = {https://doi.org/10.5281/zenodo.14845712}
}

@Article{Read:2002hq,
    Author = {Read, Alexander L.},
    Journal = {J. Phys. G},
    Pages = {2693},
    doi = "10.1088/0954-3899/28/10/313",
    Title = "{Presentation of search results: the \(CL_s\) technique}",
    Volume = {28},
    Year = {2002}
}

@article{Junk:1999kv,
    author = "Junk, Thomas",
    title = "{Confidence level computation for combining searches with small statistics}",
    eprint = "hep-ex/9902006",
    archivePrefix = "arXiv",
    reportNumber = "CARLETON-OPAL-PHYS-99-01, CERN-EP-99-041",
    doi = "10.1016/S0168-9002(99)00498-2",
    journal = "Nucl. Instrum. Meth. A",
    volume = "434",
    pages = "435--443",
    year = "1999"
}

@Article{Cowan:2010js-err,
   Author = {Cowan, Glen and Cranmer, Kyle and Gross, Eilam and Vitells, Ofer},
   Journal = {Eur. Phys. J. C},
   Volume = {73},
   Year = {2013},
   Pages = {2501},
   doi = "10.1140/epjc/s10052-013-2501-z",
}

@article{Rolke:2004mj,
    author = "Rolke, Wolfgang A. and L{\'o}pez, Angel M. and Conrad, Jan",
    editor = "Lyons, L. and Karagoz, Muge",
    title = "{Limits and confidence intervals in the presence of nuisance parameters}",
    eprint = "physics/0403059",
    archivePrefix = "arXiv",
    doi = "10.1016/j.nima.2005.05.068",
    journal = "Nucl. Instrum. Meth. A",
    volume = "551",
    pages = "493--503",
    year = "2005"
}

@Article{PERF-2007-01,
    author         = "{ATLAS Collaboration}",
    title          = "{The ATLAS Experiment at the CERN Large Hadron Collider}",
    journal        = "JINST",
    volume         = "3",
    year           = "2008",
    pages          = "S08003",
    doi            = "10.1088/1748-0221/3/08/S08003",
    primaryClass   = "hep-ex",
}

@Article{DAPR-2010-01,
    author         = "{ATLAS Collaboration}",
    title          = "{Luminosity determination in \(pp\) collisions at \(\sqrt{s} = 7\,\text{TeV}\) using the ATLAS detector at the LHC}",
    journal        = "Eur. Phys. J. C",
    volume         = "71",
    year           = "2011",
    pages          = "1630",
    doi            = "10.1140/epjc/s10052-011-1630-5",
    reportNumber   = "CERN-PH-EP-2010-069",
    eprint         = "1101.2185",
    archivePrefix  = "arXiv",
    primaryClass   = "hep-ex",
}

@Article{SOFT-2010-01,
    author         = "{ATLAS Collaboration}",
    title          = "{The ATLAS Simulation Infrastructure}",
    journal        = "Eur. Phys. J. C",
    volume         = "70",
    year           = "2010",
    pages          = "823",
    doi            = "10.1140/epjc/s10052-010-1429-9",
    eprint         = "1005.4568",
    archivePrefix  = "arXiv",
    primaryClass   = "physics.ins-det",
}

@Article{TOPQ-2012-20,
    author         = "{ATLAS Collaboration}",
    title          = "{Measurement of the production cross-section of a single top quark in association with a \(W\) boson at \(8\,\text{TeV}\) with the ATLAS experiment}",
    journal        = "JHEP",
    volume         = "01",
    year           = "2016",
    pages          = "064",
    doi            = "10.1007/JHEP01(2016)064",
    reportNumber   = "CERN-PH-EP-2015-230",
    eprint         = "1510.03752",
    archivePrefix  = "arXiv",
    primaryClass   = "hep-ex",
}

@Article{PERF-2014-03,
    author         = "{ATLAS Collaboration}",
    title          = "{Performance of pile-up mitigation techniques for jets in \(pp\) collisions at \(\sqrt{s} = 8\,\text{TeV}\) using the ATLAS detector}",
    journal        = "Eur. Phys. J. C",
    volume         = "76",
    year           = "2016",
    pages          = "581",
    doi            = "10.1140/epjc/s10052-016-4395-z",
    reportNumber   = "CERN-PH-EP-2015-206",
    eprint         = "1510.03823",
    archivePrefix  = "arXiv",
    primaryClass   = "hep-ex",
}

@Article{PERF-2015-09,
    author         = "{ATLAS Collaboration}",
    title          = "{Jet reconstruction and performance using particle flow with the ATLAS Detector}",
    journal        = "Eur. Phys. J. C",
    volume         = "77",
    year           = "2017",
    pages          = "466",
    doi            = "10.1140/epjc/s10052-017-5031-2",
    reportNumber   = "CERN-EP-2017-024",
    eprint         = "1703.10485",
    archivePrefix  = "arXiv",
    primaryClass   = "hep-ex",
}

@Article{PERF-2015-10,
    author         = "{ATLAS Collaboration}",
    title          = "{Muon reconstruction performance of the ATLAS detector in proton--proton collision data at \(\sqrt{s} = 13\,\text{TeV}\)}",
    journal        = "Eur. Phys. J. C",
    volume         = "76",
    year           = "2016",
    pages          = "292",
    doi            = "10.1140/epjc/s10052-016-4120-y",
    reportNumber   = "CERN-EP-2016-033",
    eprint         = "1603.05598",
    archivePrefix  = "arXiv",
    primaryClass   = "hep-ex",
}

@Article{TOPQ-2016-11,
    author         = "{ATLAS Collaboration}",
    title          = "{Measurement of the \(t\bar{t}Z\) and \(t\bar{t}W\) cross sections in proton--proton collisions at \(\sqrt{s} = 13\,\text{TeV}\) with the ATLAS detector}",
    journal        = "Phys. Rev. D",
    volume         = "99",
    year           = "2019",
    pages          = "072009",
    doi            = "10.1103/PhysRevD.99.072009",
    reportNumber   = "CERN-EP-2018-331",
    eprint         = "1901.03584",
    archivePrefix  = "arXiv",
    primaryClass   = "hep-ex",
}

@Article{TRIG-2016-01,
    author         = "{ATLAS Collaboration}",
    title          = "{Performance of the ATLAS trigger system in 2015}",
    journal        = "Eur. Phys. J. C",
    volume         = "77",
    year           = "2017",
    pages          = "317",
    doi            = "10.1140/epjc/s10052-017-4852-3",
    reportNumber   = "CERN-EP-2016-241",
    eprint         = "1611.09661",
    archivePrefix  = "arXiv",
    primaryClass   = "hep-ex",
}

@Article{PERF-2017-01,
    author         = "{ATLAS Collaboration}",
    title          = "{Electron reconstruction and identification in the ATLAS experiment using the 2015 and 2016 LHC proton--proton collision data at \(\sqrt{s} = 13\,\text{TeV}\)}",
    journal        = "Eur. Phys. J. C",
    volume         = "79",
    year           = "2019",
    pages          = "639",
    doi            = "10.1140/epjc/s10052-019-7140-6",
    reportNumber   = "CERN-EP-2018-273",
    eprint         = "1902.04655",
    archivePrefix  = "arXiv",
    primaryClass   = "physics.ins-det",
}

@Article{TOPQ-2017-12,
    author         = "{ATLAS Collaboration}",
    title          = "{Measurements of inclusive and differential fiducial cross-sections of \(t\bar{t}\) production with additional heavy-flavour jets in proton--proton collisions at \(\sqrt{s} = 13\,\text{TeV}\) with the ATLAS detector}",
    journal        = "JHEP",
    volume         = "04",
    year           = "2019",
    pages          = "046",
    doi            = "10.1007/JHEP04(2019)046",
    reportNumber   = "CERN-EP-2018-276",
    eprint         = "1811.12113",
    archivePrefix  = "arXiv",
    primaryClass   = "hep-ex",
}

@Article{DAPR-2018-01,
    author         = "{ATLAS Collaboration}",
    title          = "{ATLAS data quality operations and performance for 2015--2018 data-taking}",
    journal        = "JINST",
    volume         = "15",
    year           = "2020",
    pages          = "P04003",
    doi            = "10.1088/1748-0221/15/04/P04003",
    reportNumber   = "CERN-EP-2019-207",
    eprint         = "1911.04632",
    archivePrefix  = "arXiv",
    primaryClass   = "physics.ins-det",
}

@Article{EGAM-2018-01,
    author         = "{ATLAS Collaboration}",
    title          = "{Electron and photon performance measurements with the ATLAS detector using the 2015--2017 LHC proton--proton collision data}",
    journal        = "JINST",
    volume         = "14",
    year           = "2019",
    pages          = "P12006",
    doi            = "10.1088/1748-0221/14/12/P12006",
    reportNumber   = "CERN-EP-2019-145",
    eprint         = "1908.00005",
    archivePrefix  = "arXiv",
    primaryClass   = "hep-ex",
}

@Article{EXOT-2018-49,
    author         = "{ATLAS Collaboration}",
    title          = "{Search for \(t\bar{t}\) resonances in final states with exactly one or two leptons using \(140\,\text{fb}^{-1}\) of \(pp\) collision data at \(\sqrt{s}=13\,\text{TeV}\) with the ATLAS experiment}",
    year           = "2025",
    reportNumber   = "CERN-EP-2025-281",
    eprint         = "2512.17856",
    archivePrefix  = "arXiv",
    primaryClass   = "hep-ex",
}

@Article{FTAG-2018-01,
    author         = "{ATLAS Collaboration}",
    title          = "{ATLAS \(b\)-jet identification performance and efficiency measurement with \(t\bar{t}\) events in \(pp\) collisions at \(\sqrt{s} = 13\,\text{TeV}\)}",
    journal        = "Eur. Phys. J. C",
    volume         = "79",
    year           = "2019",
    pages          = "970",
    doi            = "10.1140/epjc/s10052-019-7450-8",
    reportNumber   = "CERN-EP-2019-132",
    eprint         = "1907.05120",
    archivePrefix  = "arXiv",
    primaryClass   = "hep-ex",
}

@Article{JETM-2018-05,
    author         = "{ATLAS Collaboration}",
    title          = "{Jet energy scale and resolution measured in proton--proton collisions at \(\sqrt{s} = 13\,\text{TeV}\) with the ATLAS detector}",
    journal        = "Eur. Phys. J. C",
    volume         = "81",
    year           = "2021",
    pages          = "689",
    doi            = "10.1140/epjc/s10052-021-09402-3",
    reportNumber   = "CERN-EP-2020-083",
    eprint         = "2007.02645",
    archivePrefix  = "arXiv",
    primaryClass   = "hep-ex",
}

@Article{MUON-2018-03,
    author         = "{ATLAS Collaboration}",
    title          = "{Muon reconstruction and identification efficiency in ATLAS using the full Run~2 \(pp\) collision data set at \(\sqrt{s} = 13\,\text{TeV}\)}",
    journal        = "Eur. Phys. J. C",
    volume         = "81",
    year           = "2021",
    pages          = "578",
    doi            = "10.1140/epjc/s10052-021-09233-2",
    reportNumber   = "CERN-EP-2020-199",
    eprint         = "2012.00578",
    archivePrefix  = "arXiv",
    primaryClass   = "hep-ex",
}

@Article{STDM-2018-03,
    author         = "{ATLAS Collaboration}",
    title          = "{Measurement of \(W^{\pm}Z\) production cross sections and gauge boson polarisation in \(pp\) collisions at \(\sqrt{s} = 13\,\text{TeV}\) with the ATLAS detector}",
    journal        = "Eur. Phys. J. C",
    volume         = "79",
    year           = "2019",
    pages          = "535",
    doi            = "10.1140/epjc/s10052-019-7027-6",
    reportNumber   = "CERN-EP-2018-327",
    eprint         = "1902.05759",
    archivePrefix  = "arXiv",
    primaryClass   = "hep-ex",
}

@Article{TOPQ-2018-05,
    author         = "{ATLAS Collaboration}",
    title          = "{Evidence for \(t\bar{t}t\bar{t}\) production in the multilepton final state in proton--proton collisions at \(\sqrt{s} = 13\text {TeV}\) with the ATLAS detector}",
    journal        = "Eur. Phys. J. C",
    volume         = "80",
    year           = "2020",
    pages          = "1085",
    doi            = "10.1140/epjc/s10052-020-08509-3",
    reportNumber   = "CERN-EP-2020-111",
    eprint         = "2007.14858",
    archivePrefix  = "arXiv",
    primaryClass   = "hep-ex",
}

@Article{TRIG-2018-01,
    author         = "{ATLAS Collaboration}",
    title          = "{Performance of the ATLAS muon triggers in Run~2}",
    journal        = "JINST",
    volume         = "15",
    year           = "2020",
    pages          = "P09015",
    doi            = "10.1088/1748-0221/15/09/p09015",
    reportNumber   = "CERN-EP-2020-031",
    eprint         = "2004.13447",
    archivePrefix  = "arXiv",
    primaryClass   = "physics.ins-det",
}

@Article{TRIG-2018-05,
    author         = "{ATLAS Collaboration}",
    title          = "{Performance of electron and photon triggers in ATLAS during LHC Run~2}",
    journal        = "Eur. Phys. J. C",
    volume         = "80",
    year           = "2020",
    pages          = "47",
    doi            = "10.1140/epjc/s10052-019-7500-2",
    reportNumber   = "CERN-EP-2019-169",
    eprint         = "1909.00761",
    archivePrefix  = "arXiv",
    primaryClass   = "hep-ex",
}

@Article{FTAG-2019-02,
    author         = "{ATLAS Collaboration}",
    title          = "{Calibration of the light-flavour jet mistagging efficiency of the \(b\)-tagging algorithms with \(Z\)+jets events using \(139\,\text{fb}^{-1}\) of ATLAS proton--proton collision data at \(\sqrt{s} = 13\,\text{TeV}\)}",
    journal        = "Eur. Phys. J. C",
    volume         = "83",
    year           = "2023",
    pages          = "728",
    doi            = "10.1140/epjc/s10052-023-11736-z",
    reportNumber   = "CERN-EP-2022-211",
    eprint         = "2301.06319",
    archivePrefix  = "arXiv",
    primaryClass   = "hep-ex",
}

@Article{GENR-2019-02,
    author         = "{ATLAS Collaboration}",
    title          = "{The ATLAS experiment at the CERN Large Hadron Collider: a description of the detector configuration for Run~3}",
    journal        = "JINST",
    volume         = "19",
    year           = "2024",
    pages          = "P05063",
    doi            = "10.1088/1748-0221/19/05/P05063",
    reportNumber   = "CERN-EP-2022-259",
    eprint         = "2305.16623",
    archivePrefix  = "arXiv",
    primaryClass   = "physics.ins-det",
}

@Article{FTAG-2020-08,
    author         = "{ATLAS Collaboration}",
    title          = "{Measurement of the \(c\)-jet mistagging efficiency in \(t\bar{t}\) events using \(pp\) collision data at \(\sqrt{s} = 13\,\text{TeV}\) collected with the ATLAS detector}",
    journal        = "Eur. Phys. J. C",
    volume         = "82",
    year           = "2022",
    pages          = "95",
    doi            = "10.1140/epjc/s10052-021-09843-w",
    reportNumber   = "CERN-EP-2021-166",
    eprint         = "2109.10627",
    archivePrefix  = "arXiv",
    primaryClass   = "hep-ex",
}

@Article{JETM-2020-03,
    author         = "{ATLAS Collaboration}",
    title          = "{The performance of missing transverse momentum reconstruction and its significance with the ATLAS detector using \(140\,\text{fb}^{-1}\) of \(\sqrt{s} = 13\,\text{TeV}\) \(pp\) collisions}",
    journal        = "Eur. Phys. J. C",
    volume         = "85",
    year           = "2025",
    pages          = "606",
    doi            = "10.1140/epjc/s10052-025-14062-8",
    reportNumber   = "CERN-EP-2024-023",
    eprint         = "2402.05858",
    archivePrefix  = "arXiv",
    primaryClass   = "hep-ex",
}

@Article{DAPR-2021-01,
    author         = "{ATLAS Collaboration}",
    title          = "{Luminosity determination in \(pp\) collisions at \(\sqrt{s} = 13\,\text{TeV}\) using the ATLAS detector at the LHC}",
    journal        = "Eur. Phys. J. C",
    volume         = "83",
    year           = "2023",
    pages          = "982",
    doi            = "10.1140/epjc/s10052-023-11747-w",
    reportNumber   = "CERN-EP-2022-281",
    eprint         = "2212.09379",
    archivePrefix  = "arXiv",
    primaryClass   = "hep-ex",
}

@Article{EGAM-2021-01,
    author         = "{ATLAS Collaboration}",
    title          = "{Electron and photon efficiencies in LHC Run~2 with the ATLAS experiment}",
    journal        = "JHEP",
    volume         = "05",
    year           = "2024",
    pages          = "162",
    doi            = "10.1007/JHEP05(2024)162",
    reportNumber   = "CERN-EP-2023-182",
    eprint         = "2308.13362",
    archivePrefix  = "arXiv",
    primaryClass   = "hep-ex",
}

@Article{EGAM-2021-02,
    author         = "{ATLAS Collaboration}",
    title          = "{Electron and photon energy calibration with the ATLAS detector using LHC Run~2 data}",
    journal        = "JINST",
    volume         = "19",
    year           = "2024",
    pages          = "P02009",
    doi            = "10.1088/1748-0221/19/02/P02009",
    reportNumber   = "CERN-EP-2023-128",
    eprint         = "2309.05471",
    archivePrefix  = "arXiv",
    primaryClass   = "hep-ex",
}

@Article{IDTR-2021-01,
    author         = "{ATLAS Collaboration}",
    title          = "{Track and Vertex Reconstruction with the ATLAS Inner Detector}",
    year           = "2026",
    reportNumber   = "CERN-EP-2026-096",
    eprint         = "2605.07585",
    archivePrefix  = "arXiv",
    primaryClass   = "physics.ins-det",
}

@Article{EXOT-2022-14,
    author         = "{ATLAS Collaboration}",
    title          = "{Search for top-philic heavy resonances in \(pp\) collisions at \(\sqrt{s} = 13\,\text{TeV}\) with the ATLAS detector}",
    journal        = "Eur. Phys. J. C",
    volume         = "84",
    year           = "2024",
    pages          = "157",
    doi            = "10.1140/epjc/s10052-023-12318-9",
    reportNumber   = "CERN-EP-2023-048",
    eprint         = "2304.01678",
    archivePrefix  = "arXiv",
    primaryClass   = "hep-ex",
}

@Article{MUON-2022-01,
    author         = "{ATLAS Collaboration}",
    title          = "{Studies of the muon momentum calibration and performance of the ATLAS detector with \(pp\) collisions at \(\sqrt{s} = 13\,\text{TeV}\)}",
    journal        = "Eur. Phys. J. C",
    volume         = "83",
    year           = "2023",
    pages          = "686",
    doi            = "10.1140/epjc/s10052-023-11584-x",
    reportNumber   = "CERN-EP-2022-274",
    eprint         = "2212.07338",
    archivePrefix  = "arXiv",
    primaryClass   = "hep-ex",
}

@Article{TRIG-2022-01,
    author         = "{ATLAS Collaboration}",
    title          = "{The ATLAS trigger system for LHC Run~3 and trigger performance in 2022}",
    journal        = "JINST",
    volume         = "19",
    year           = "2024",
    pages          = "P06029",
    doi            = "10.1088/1748-0221/19/06/P06029",
    reportNumber   = "CERN-EP-2023-299",
    eprint         = "2401.06630",
    archivePrefix  = "arXiv",
    primaryClass   = "hep-ex",
}

@Article{FTAG-2023-05,
    author         = "{ATLAS Collaboration}",
    title          = "{Transforming jet flavour tagging at ATLAS}",
    journal        = "Nature Commun.",
    volume         = "17",
    year           = "2026",
    pages          = "541",
    doi            = "10.1038/s41467-025-65059-6",
    reportNumber   = "CERN-EP-2025-103",
    eprint         = "2505.19689",
    archivePrefix  = "arXiv",
    primaryClass   = "hep-ex",
}

@Article{FTAG-2023-10,
    author         = "{ATLAS Collaboration}",
    title          = "{Measurement of the \(b\)-jet identification efficiency in dileptonic \(t\bar{t}\) events using proton--proton collision data at \(\sqrt{s} = 13.6\,\text{TeV}\) collected with the ATLAS detector}",
    year           = "2026",
    reportNumber   = "CERN-EP-2026-156",
    eprint         = "2607.05322",
    archivePrefix  = "arXiv",
    primaryClass   = "hep-ex",
}

@Article{TOPQ-2023-19,
    author         = "{ATLAS Collaboration}",
    title          = "{Climbing to the Top of the ATLAS \(13\,\text{TeV}\) data}",
    journal        = "Phys. Rept.",
    volume         = "1116",
    year           = "2024",
    pages          = "127--183",
    doi            = "10.1016/j.physrep.2024.12.004",
    reportNumber   = "CERN-EP-2024-099",
    eprint         = "2404.10674",
    archivePrefix  = "arXiv",
    primaryClass   = "hep-ex",
}

@Article{FTAG-2024-07,
    author         = "{ATLAS Collaboration}",
    title          = "{Simultaneous efficiency measurements of \(b\)- and \(c\)-jets in \(t\bar{t}\) events from \(\sqrt{s} = 13.6\,\text{TeV}\) \(pp\) collision data collected with the ATLAS detector}",
    year           = "2026",
    reportNumber   = "CERN-EP-2025-292",
    eprint         = "2607.06199",
    archivePrefix  = "arXiv",
    primaryClass   = "hep-ex",
}

@Article{EXOT-2025-09,
    author         = "{ATLAS Collaboration}",
    title          = "{Search for high-mass dilepton resonances in \(pp\) collisions at \(\sqrt{s} = 13.6\,\text{TeV}\) combined with \(13\,\text{TeV}\) results using the ATLAS detector}",
    year           = "2026",
    reportNumber   = "CERN-EP-2026-230",
    eprint         = "2607.28334",
    archivePrefix  = "arXiv",
    primaryClass   = "hep-ex",
}

@Booklet{ATL-SOFT-PUB-2021-001,
    author         = "{ATLAS Collaboration}",
    title          = "{The ATLAS Collaboration Software and Firmware}",
    howpublished   = "{ATL-SOFT-PUB-2021-001}",
    url            = "https://cds.cern.ch/record/2767187",
    year           = "2021",
}

@Booklet{ATL-SOFT-PUB-2026-001,
    author         = "{ATLAS Collaboration}",
    title          = "{ATLAS Computing Acknowledgements}",
    howpublished   = "{ATL-SOFT-PUB-2026-001}",
    url            = "https://cds.cern.ch/record/2952666",
    year           = "2026",
}

@Article{CMS-B2G-17-017,
    author         = "{CMS Collaboration}",
    title          = "{Search for resonant \(t\bar{t}\) production in proton--proton collisions at \(\sqrt{s} = 13\,\text{TeV}\)}",
    journal        = "JHEP",
    volume         = "04",
    year           = "2019",
    pages          = "031",
    doi            = "10.1007/JHEP04(2019)031",
    reportNumber   = "CERN-EP-2018-254",
    eprint         = "1810.05905",
    archivePrefix  = "arXiv",
    primaryClass   = "hep-ex",
}

@Article{CMS-EXO-19-019,
    author         = "{CMS Collaboration}",
    title          = "{Search for resonant and nonresonant new phenomena in high-mass dilepton final states at \(\sqrt{s} = 13\,\text{TeV}\)}",
    journal        = "JHEP",
    volume         = "07",
    year           = "2021",
    pages          = "208",
    doi            = "10.1007/JHEP07(2021)208",
    reportNumber   = "CERN-EP-2021-026",
    eprint         = "2103.02708",
    archivePrefix  = "arXiv",
    primaryClass   = "hep-ex",
}

@Article{CMS-TOP-24-008,
    author         = "{CMS Collaboration}",
    title          = "{Search for physics beyond the standard model in four and three top quark production events using proton--proton collisions at \(\sqrt{s} = 13\,\text{TeV}\)}",
    year           = "2026",
    reportNumber   = "CERN-EP-2026-140",
    eprint         = "2606.30862",
    archivePrefix  = "arXiv",
    primaryClass   = "hep-ex",
}

@Booklet{ATL-PHYS-PUB-2015-026,
    author         = "{ATLAS Collaboration}",
    title          = "{Vertex Reconstruction Performance of the ATLAS Detector at \(\sqrt{s} = 13\,\text{TeV}\)}",
    howpublished   = "{ATL-PHYS-PUB-2015-026}",
    url            = "https://cds.cern.ch/record/2037717",
    year           = "2015",
}

@Booklet{ATL-PHYS-PUB-2016-020,
    author         = "{ATLAS Collaboration}",
    title          = "{Studies on top-quark Monte Carlo modelling for Top2016}",
    howpublished   = "{ATL-PHYS-PUB-2016-020}",
    url            = "https://cds.cern.ch/record/2216168",
    year           = "2016",
}

@Booklet{ATL-DAPR-PUB-2024-001,
    author         = "{ATLAS Collaboration}",
    title          = "{Preliminary analysis of the luminosity calibration for the ATLAS \(13.6\,\text{TeV}\) data recorded in 2023}",
    howpublished   = "{ATL-DAPR-PUB-2024-001}",
    url            = "https://cds.cern.ch/record/2900949",
    year           = "2024",
}
\clearpage
% ATLAS Collaboration author list
% Reference date of EXOT-2022-44 is 2026-07-17
% Author list last updated on date 26-AUG-26
% Data extracted on 26-Aug-2026 for paper reference EXOT-2022-44
% at 1:15pm
 
\begin{flushleft}
\hypersetup{urlcolor=black}
{\Large The ATLAS Collaboration}

\bigskip

\AtlasOrcid[0000-0002-6665-4934]{G.~Aad}$^\textrm{\scriptsize 102}$,
\AtlasOrcid[0000-0001-7616-1554]{E.~Aakvaag}$^\textrm{\scriptsize 17}$,
\AtlasOrcid[0000-0002-5888-2734]{B.~Abbott}$^\textrm{\scriptsize 121}$,
\AtlasOrcid[0000-0002-0287-5869]{S.~Abdelhameed}$^\textrm{\scriptsize 83b}$,
\AtlasOrcid[0000-0002-1002-1652]{K.~Abeling}$^\textrm{\scriptsize 54}$,
\AtlasOrcid[0000-0001-5763-2760]{N.J.~Abicht}$^\textrm{\scriptsize 48}$,
\AtlasOrcid[0000-0002-8496-9294]{S.H.~Abidi}$^\textrm{\scriptsize 30}$,
\AtlasOrcid[0009-0003-6578-220X]{M.~Aboelela}$^\textrm{\scriptsize 44}$,
\AtlasOrcid[0000-0002-9987-2292]{A.~Aboulhorma}$^\textrm{\scriptsize 36e}$,
\AtlasOrcid[0000-0001-5329-6640]{H.~Abramowicz}$^\textrm{\scriptsize 154}$
% ...
\bigskip
\\

% ...
$^{17}$Department for Physics and Technology, University of Bergen, Bergen; Norway.\\
% ...
$^{30}$Physics Department, Brookhaven National Laboratory, Upton NY; United States of America.\\
% ...
$^{36}$$^{(a)}$Facult\'e des Sciences Ain Chock, Universit\'e Hassan II de Casablanca;$^{(b)}$Facult\'{e} des Sciences, Universit\'{e} Ibn-Tofail, K\'{e}nitra;$^{(c)}$Facult\'e des Sciences Semlalia, Universit\'e Cadi Ayyad, LPHEA-Marrakech;$^{(d)}$LPMR, Facult\'e des Sciences, Universit\'e Mohamed Premier, Oujda;$^{(e)}$Facult\'e des sciences, Universit\'e Mohammed V, Rabat;$^{(f)}$Institute of Applied Physics, Mohammed VI Polytechnic University, Ben Guerir; Morocco.\\
% ...
$^{44}$Physics Department, Southern Methodist University, Dallas TX; United States of America.\\
% ...
$^{48}$Fakult\"{a}t Physik, Technische Universit{\"a}t Dortmund, Dortmund; Germany.\\
% ...
$^{54}$II. Physikalisches Institut, Georg-August-Universit\"{a}t G\"ottingen, G\"ottingen; Germany.\\
% ...
$^{83}$$^{(a)}$Khalifa University of Science and Technology, Abu Dhabi;$^{(b)}$New York University Abu Dhabi, Abu Dhabi;$^{(c)}$United Arab Emirates University, Al Ain;$^{(d)}$University of Sharjah, Sharjah; United Arab Emirates.\\
% ...
$^{102}$CPPM, Aix-Marseille Universit\'e, CNRS/IN2P3, Marseille; France.\\
% ...
$^{121}$Homer L. Dodge Department of Physics and Astronomy, University of Oklahoma, Norman OK; United States of America.\\
% ...
$^{154}$Raymond and Beverly Sackler School of Physics and Astronomy, Tel Aviv University, Tel Aviv; Israel.\\
% ...
$^{a}$ Also at Affiliated with an institute formerly covered by a cooperation agreement with CERN.\\
$^{b}$ Also at An-Najah National University, Nablus; Palestine.\\
$^{c}$ Also at Borough of Manhattan Community College, City University of New York, New York NY; United States of America.\\
$^{d}$ Also at Center for Interdisciplinary Research and Innovation (CIRI-AUTH), Thessaloniki; Greece.\\
$^{e}$ Also at Centre of Physics of the Universities of Minho and Porto (CF-UM-UP); Portugal.\\
$^{f}$ Also at CERN, Geneva; Switzerland.\\
% ...

\end{flushleft}

% Created with Glance <Atlas.Glance@cern.ch>

%

%
%

%

%
%
%
%

\end{document}